\documentclass[fleqn,usenatbib]{mnras}

\usepackage{newtxtext,newtxmath}
\usepackage[T1]{fontenc}

\DeclareRobustCommand{\VAN}[3]{#2}
\let\VANthebibliography\thebibliography
\def\thebibliography{\DeclareRobustCommand{\VAN}[3]{##3}\VANthebibliography}

\usepackage{graphicx}	% Including figure files
\usepackage{amsmath}	% Advanced maths commands
\usepackage{comment}
\usepackage{bm} 
\usepackage{multicol}

\title[CMZ ionization from 511 keV sources]{Do the sources of the 511 keV excess explain the anomalous CMZ ionization?} %{$\beta^+$ radionuclides explaining the 511 keV excess solve the anomalous ionization in the CMZ}

\author[De la Torre Luque, P. \& Calore F.]{Pedro De la Torre Luque$^{1, \,2}$\thanks{E-mail: pedro.delatorre@uam.es}
Francesca Calore,$^{3}$ \thanks{E-mail: francesca.calore@lapth.fr}
\\
$^{1}$Departamento de F\'{i}sica Te\'{o}rica, M-15, Universidad Aut\'{o}noma de Madrid, E-28049 Madrid, Spain\\
$^{2}$Instituto de F\'{i}sica Te\'{o}rica UAM-CSIC, Universidad Aut\'{o}noma de Madrid, C/ Nicol\'{a}s Cabrera, 13-15, 28049 Madrid, Spain\\
$^{3}$Another Department, Different Institution, Street Address, City Postal Code, Country
}

\date{Accepted XXX. Received YYY; in original form ZZZ}

\pubyear{\the\year{}}

\begin{document}
\label{firstpage}
\pagerange{\pageref{firstpage}--\pageref{lastpage}}
\maketitle

% Abstract of the paper
\begin{abstract}
The anomalous rate of molecules ionization observed at the Central Molecular zone (CMZ) challenges known mechanisms of ionization observed in molecular clouds across the Galaxy, due to the exceptionally high levels of ionization measured (orders of magnitude above what cosmic rays can explain) and its uniform spatial distribution within the CMZ. %Recently, Ref.~\cite{delatorre}
Recent studies suggest that the source of the $511$~keV excess can be correlated with this anomalous ionization rate or contribute significantly to the ionization in the Galactic Centre (GC). 
One of the leading hypotheses attributes the $511$~keV signal to positron injection from radionuclides or pulsars distributed following the stellar bulge, which is rather flat around the GC and, hence, could help explaining the uniform ionization profile.
%In fact, one of the leading explanations for the $511$~keV excess is the injection of positrons from radionuclides or pulsars following the bulge distribution of stars, which is rather flat at the center GC, which would help in producing an uniform distribution of the ionization rate. 
In this work, we investigate whether such a population of sources, injecting MeV positrons at rates consistent with the $511$~keV observations, can account for the ionization levels and distribution observed in the CMZ. Our results indicate that positron injection alone falls short at explaining the anomaly, although their expected ionization is larger than expected from any previously studied candidates.%, and we discuss the implications and viability of such population of sources as the origin of the observed excess ionization.
\end{abstract}

% Select between one and six entries from the list of approved keywords.
% Don't make up new ones.

%%%%%%%%%%%%%%%%%%%%%%%%%%%%%%%%%%%%%%%%%%%%%%%%%%

\section{Introduction}
Observations of the $511$~keV line emission have been long pursued to study sources of positrons in the Galaxy~\citep{Johnson1972ApJ172L1, Haymes1975ApJ201593, Leventhal1978ApJ225L11, Siegert2023PositronPuzzle, Prantzos_2011}. Positrons injected in the Galaxy interact with the interstellar medium (ISM) and lose energy through different mechanisms, until they either leave the Galaxy (the most energetic ones) or become thermal~\citep{Guessoum:2005cb, Guessoum1991}. Once they thermalize, they have a high chance of encountering with ISM electrons (free electrons, electrons bounded to molecules or atoms or electrons in dust particles) and will either annihilate directly or form a positronium bound state~\citep{Stecker, Guessoum:2005cb}. The latter process is more efficient and, especially, the cross section of charge exchange (e$^-$ $+$ H$_2$ $\longrightarrow$ ps + H$_2^-$) is more than $10^6$ times larger than the Dirac e$^+$e$^-$ direct annihilation cross sections~\citep{JeanP_2009}. Therefore, it is expected that the $511$~keV emission is related not only to the positrons injected but also to the medium where they become thermal and its gas distribution.

After decades of observations, we still do not understand the very high intensity of $511$~keV photons observed around the center of the Galaxy - the so-called ``positron puzzle'', which requires the injection of a large flux of positrons, at an estimated rate around a few times $10^{43}$~e$^+$/s~\citep{Siegert2016AandA586A84, AharonianAtoyan1981SovAstrLett, kierans2019positronannihilationgalaxy}. Different sources have been proposed to explain the morphology and intensity of the signal. We list below some of the most popular candidates~\citep{Prantzos_2011}: 
\begin{itemize}
    \item The radioactive decay of $\beta^+$ unstable isotopes formed in different systems %($^{56}$Co, $^{44}$Ti, $^{26}$Al, or $^{22}$Na)
     is a guaranteed source of positrons in the Galaxy~\citep{clayton1973positronium, AharonianAtoyan1981SovAstrLett, ChanLingenfelter1993ApJ405614}. %Stellar nucleosynthesis products were proposed as a primary source of Galactic positrons long time ago~\citep{clayton1973positronium, AharonianAtoyan1981SovAstrLett, ChanLingenfelter1993ApJ405614}. 
    Since massive stars should be more concentrated around the Galactic center (GC), these sources received special attention as potential explanation of the $511$~keV excess~\citep{Diehl1995AandA298445, Knoedlseder2005AandA441513, Weidenspointner2006AandA4501013, Siegert2023PositronPuzzle}.
    In particular, the production of $^{26}$Al from massive stars is expected to be an important fraction (between $10\%$ and $100\%$) of the $511$~keV photons observed in the Galactic disk~\citep{Prantzos_2011, Skinner2015Integral2014_511keV, Siegert2022MNRAS509L11}.
    On top of that, another guaranteed contribution must come from radionuclides produced in supernovae (SNe) (as $^{44}$Ti in core-collapse SNe or $^{56}$Co -- produced in type IA SNe), or $^{22}$Na. %Another possible contributor is $^{22}$Na, with a short lifetime of $2.6$~years, that is also produced in SNe and novae. %Unfortunately, the nova and SN explosion rates in the vicinity of the GC remain poorly constrained, along with the star formation rate and the expected population of massive stars. Besides these events happening all around the disk, they are expected to be relevantly concentrated following the observed bulge distribution, therefore producing a sea of positrons that would be expected to follow bulge stars. All these sources would produce near MeV positrons following the $\beta^+$-decay spectrum (see Sec.~\ref{sec:RadPL} for details).
%Type Ia SNe (SN Ia) are often considered as the most plausible source of positrons in the Milky Way (Dermer \& Murphy 2001). SN Ia produce positrons via the beta+-decay of radioactive 56Co (tau =111 days). Expected 56Co yields of $0.6$~M$_\circ$ provide $\sim 2.5\cdot10^{54}$ positrons per event, although, as with novae, prompt annihilation in the SN envelope probably prevents large fractions of the positrons  from escaping into the ISM. From the analysis of late light curves of SN Ia Milne et al. (1999) derive a mean escaped positron yield of $\sim 8 \cdot 10^{52}$ positrons per SN Ia, corresponding to a positron escape fraction of f $\sim$ 0.03. A recent study of SN 2000cx even suggests f - 0, but SN 2000cx was an unusual event that may not represent the average SN in the bulge of our Galaxy (Sollerman et al. 2004). \textcolor{red}{from arXiv:astro-ph/0506026}
    \item Accreting binary systems are also expected to produce positrons and accelerate them~\citep{Paredes2005, Li1996, Guessoum_2006, bandyopadhyay2009origin}. Microquasars have long been proposed as a source of Galactic positrons -- from the potential detection of $511$~keV photons from the “Great Annihilator” by SIGMA~\citep{Mirabel1992} (but later refuted by \cite{Smith1996}). %A similar case was that of the microquasar V404 Cygni~\citep{Siegert2016}, which was also controversial~\citep{Roques2016}. %In microquasars, positrons can be created through different mechanisms~\citep{Paredes2005, Li1996} %pair-production in the hot inner accretion disk, in the X-ray corona, or at the base of the jet 
    %and the predicted rates of 511 keV emission from the brightest microquasars should be detectable with a next-generation instrument~\citep{Guessoum_2006}. 
    However,~\cite{Guessoum_2006}   constrained their contribution to be lower than $\sim 3 \cdot 10^{41}$e$^+$ s$^{1}$ under the assumption that they are steady positron emitters. %\textcolor{red}{Citations from arXiv:1903.05569}
    While high-mass X-ray binaries are found correlated with young stars and primarily distributed across the disk, low-mass X-ray binaries (LMXB) are strongly concentrated towards the Galactic bulge, and are promising source candidates. %Among the 150 LMXBs listed in the catalogue of Liu et al. (2001), more than 50\% are observed towards the Galactic bulge. Correcting for completeness, Grimm et al. (2002) find a B/D ratio of $\sim0.9$ and a vertical scale height of 410 pc for the LMXB distribution. 
    These sources are expected to release positrons with a spectrum that roughly follows a power law. %, however the uncertainty on their injection rates is huge, with predictions ranging from $10^{41}$ to a few times $10^{43}$~e$^+$ s$^{-1}$. 
    \item Other compact objects like pulsars are also candidates to inject positrons in the interstellar medium. The PAMELA positron excess~\citep{Adriani2009Nature458607} and, then, the discovery of TeV-halos~\citep{Abeysekara2017ApJ84340} supports predictions foreseeing these sources as high-energy positron emitters~\citep{Dan_Hooper_2009, Y_ksel_2009}. %However, although they are probably the clearest source of positrons in the ISM, along with positrons produced from cosmic-ray interactions~\citep{DeLaTorreLuque2023JCAP10_011, Di_Mauro_2023}, they are traditionally not expected to be especially concentrated around the bulge. Nevertheless, 
    In particular, millisecond pulsars are an interesting candidate for the 511 keV excess, and are still among the most promising explanation of the \textit{Fermi}-LAT GC Excess~\citep{Murgia2020ARNPS70_455}. They are expected to follow the bulge distribution and inject positrons following power-law trends. 
    \item A more exotic possibility is the one of sub-GeV dark matter annihilation~\citep{Boehm:2003bt}. Dark matter is expected to be concentrated in the inner Galaxy, following an Navarro-Frenk-White distribution~\citep{Navarro:1995iw, Navarro:2003ew} or something even more contracted\footnote{Note that current observations of the Milky Way do not allow us to robustly constrain the dark matter density in the inner Galaxy -- e.g. ~\cite{gardner2021milkyway}}. The spatial distribution of the 511 keV excess is tantalizingly similar to the expected dark matter distribution in the Milky Way from the Navarro-Frenk-White N-body simulations~\citep{Boehm:2002yz, Ascasibar_2006, DelaTorreLuque:2023cef}. Interestingly, \cite{Muru2025PRL135161005} has recently shown that the dark matter distribution can be affected by the presence of baryons in Milky Way-like Galaxies, deforming the dark matter distribution to be similar to that of the bulge. %However, from the point of view of particle physics models, common candidates are very much constrained from cosmology and might be fine tuned. Nevertheless, there are still interesting candidates compatible with current constraints~\citep{Boehm:2002yz, Cappiello_2023, Aghaie:2025dgl, balaji2025exciteddarkmattersolution}.
    From a particle physics perspective, candidates are tightly constrained by cosmology and may require fine-tuning, though viable options consistent with current limits still exist~\citep{Boehm:2002yz, Cappiello_2023, Aghaie:2025dgl, balaji2025exciteddarkmattersolution}.
    \item Accretion around the supermassive black hole at the GC (Sgr A*) can produce non-thermal emissions, including positrons~\citep{totani2006riaf}. This source has been also invoked, and there have been several attempts to find a signal in INTEGAL data~\citep{Knoedlseder2005AandA441513, Jean_2005, Siegert_2016}. However, the current limitations (mainly the poor spatial resolution of INTEGRAL) make it difficult to identify such a component.
\end{itemize}

All of these sources may potentially provide large positron injection rates; however, quantitative estimates are highly uncertain, and direct measurements of their positron injection are not accessible with current experimental capabilities. Nevertheless, their spatial distributions may offer a way to constrain their relative importance. In this regard, a template fit analysis of the $511$~keV line~\citep{Siegert2022MNRAS509L11} found that the signal favors a bulge distribution over other templates, when analyzing the signal along with the continuum ortho-positronium contribution. On top of this, the same analysis found evidence for positrons injected with low energies $E_{e^+} \lesssim 1.5$~MeV, giving further support to the radionuclides hypothesis. Such a low energy positron injection spatially associated to the $511$~keV excess has recently found further evidence from an analysis of COMPTEL data~\cite{Kn_dlseder_2025}, where the authors found a significant in-flight positron annihilation signal that is well explained with positron injection at $\sim2$~MeV (total energy). In sum, although systematic uncertainties in these analyses are still sizable, the latest analyses suggest the injection of low energy ($ \sim$MeV) positrons roughly following the bulge distribution as responsible for the 511 keV line GC signal.

Another anomaly has recently been related with the injection of MeV particles around the GC: An unexpectedly high ionization rate of gas around the GC, particularly in the Central Molecular Zone (CMZ). The CMZ is a dense molecular gas region that extends over $\sim200$~pc in radius and $\sim100$~pc high around the GC, with an average gas density of about n$_{\textrm{H}_2} \sim 150$~cm$^{-3}$. Different astrochemical tracers, and especially rovibrational lines of H$_3^+$ transitions, have revealed an unexpectedly high ionization rate of molecular hydrogen in this region, that can be as high as  $\zeta \sim 10^{-14}-10^{-15} \ \mathrm{s}^{-1}$ \citep{Oka_2005, Oka_2020, Geballe_2010, Indriolo_2009, IoRPetit, IoRMethanol, Indriolo_2015, van_der_Tak_2006}. 
This value exceeds the ionization rates measured in dense clouds 
($\zeta \sim 10^{-17}\ \mathrm{s}^{-1}$) by two to three orders of magnitude~\citep{Indriolo_2015, Phan_2018}, and surpasses the predicted ionization rates from cosmic rays by at least two orders of magnitude~\citep{Ravikularaman2025AandA694A114}. 
This may indicate the presence of a source population (or a very diffuse/extended source) highly concentrated around the GC that is emitting ionizing particles that lose their energy very quickly, via ionization of molecular gas~\citep{Indriolo_2009}. %From the latter, one can infer that these particles must be injected at low energies ($\sim$ MeV), to be efficiently confined around the CMZ thereby preventing them from propagating long distances.
The roughly uniform values of $\zeta$ measured throughout the CMZ make the anomaly even more striking, given that the main ionization candidates (particle emissions from SgrA*, point-like sources of high-energy emissions and diffuse cosmic rays) lead to strong radial dependence of the produced ionization~\citep{Dogiel_2014, Ravikularaman2025AandA694A114}. 
We also note that a recent study from \cite{Obolentseva_2024} argued that ionization rates measured in different dense molecular clouds were biased towards higher values, due to incorrect assumptions on their line-of-sight sizes and densities. However, this correction cannot explain the anomalously large ionization observed in the CMZ, whose geometry and gas properties are comparatively well measured, although it may partially lower the inferred CMZ ionization level.

Signal correlations can be extremely useful to understand the origin of this long standing excesses. In this work, we plan to test whether the emission of positrons explaining the $511$~keV signal implies a very large rate of ionization of hydrogen in the CMZ - as the one that is observed - if these positrons are produced by sources following the bulge distribution in our Galaxy.  %In particular, this explanation suggests that whichever is the source of the 511 keV emission, it will produce a ; pointing to a population of massive stars or similar sources of $\beta^{+}$ radionuclides
While the impact onto the $511$~keV signal from different gas phases has been studied in several works, the effects of the correlated positron injection on the gas surrounding the GC have never been explored. Discussing this, and in particular, estimating the ionization produced in the CMZ by the positron sources that can potentially explain the $511$~keV excess is the scope of this paper. 
To do this, we adopt different assumptions on the distribution of these sources, following the bulge structure, and the spectral dependence of the positron injection. %We consider a general case and then study what is the expected ionization from $\beta^+$-radionuclides emitted either in SNe or from massive stars, as well as pulsars/X-ray binaries. 
Finally, we discuss how relevant is such an injection of particles concentrated around the CMZ for the star formation rate and the dynamics of particle diffusion within such a structure.
%As we will show, non-thermal emission from sources distributed following the bulge distribution naturally explain the "flatness" of the ionization profile. s%Moreover, we argue that the sources explaining the $511$~keV signal could provide a natural explanation to the observed anomalous ionization rate in the CMZ. 
%Interestingly, a possible connection between the \textit{Fermi}-LAT GC excess and the CMZ anomaly was also argued in~\cite{Silk_2018}, and could indicate that these three excesses could potentially be correlated.

This paper is structured as it follows: We first describe the bulge distribution that we use to evaluate the positron source distribution in Sect.~\ref{sec:BulgeDist}. Then we discuss the positron injection spectrum and how this is normalized to the observed positron injection rate in sections~\ref{sec:PosInj} and~~\ref{sec:Normaliz}, respectively.
Then, we report the expected ionization rates and their spatial profiles in the CMZ for the different cases studied, in Section~\ref{sec:Results}. Our conclusions are summarized in Sect.~\ref{sec:Conclusions}.

\section{Methodology}
\subsection{Bulge distributions}
\label{sec:BulgeDist}
In our estimations, we employ the state-of-the-art parameterizations of the observed Galactic bulge distribution, combining a nuclear bulge, with a larger-scale bulge. 
For the latter, we use two representative cases described below. 

\begin{figure}
	\includegraphics[width=\columnwidth]{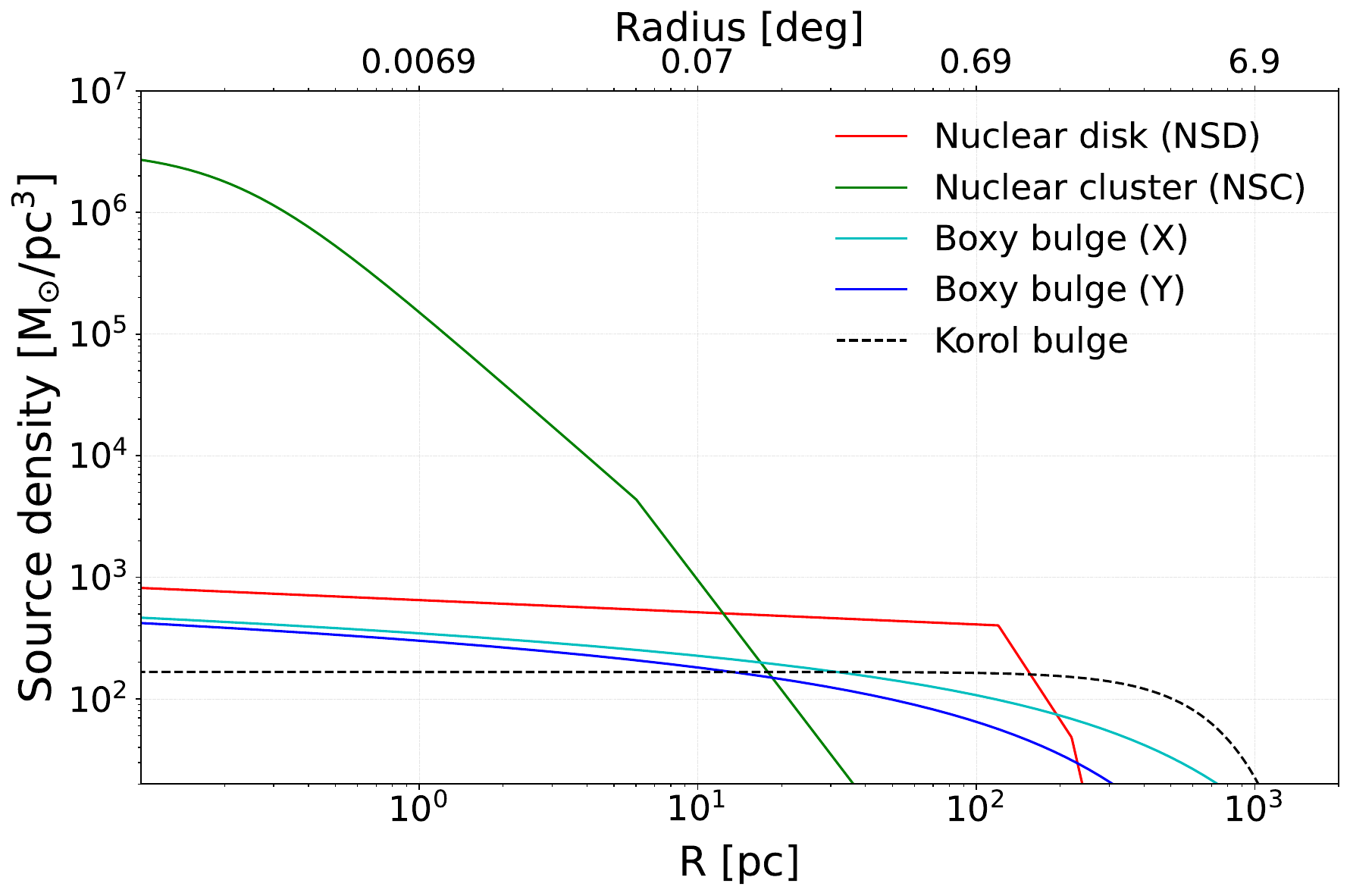}
    \caption{Comparison of the different components and models for the source distribution in the Galactic bulge, described in the text. The figure illustrates the source density as a function of radial distance for each model, with the normalizations detailed in the main text.}
    \label{fig:Distribs}
\end{figure}

\textbf{Korol distribution:} As a representative case for the bulge distribution we employ the parameterization given in~\cite{Korol_2018}, which features a radially symmetric bulge with an exponential decrease in density of stars, following:
\begin{equation}
\rho = \rho_0 \cdot \exp\left(-\frac{r^2}{2r_b^2} - \frac{z}{z_b}
\right) \, ,
\label{eq:Korol}
\end{equation}
where $r$ is the radial distance from the GC in the plane of the Galaxy and $z$ represents the vertical coordinate. $r_b$ and $z_b$ are set to $500$ and $200$~pc, respectively. As we describe later (Sec~\ref{sec:Normaliz}), the normalization density is set by imposing that the total injection rate of positrons is the one inferred from the $511$~keV observations.

\textbf{Boxy-Bulge distribution:}
The second model for the boxy-bulge follows the parameterizations fit by the triaxial E3 model explored in~\cite{Cao2013MNRAS434595, Dwek1995ApJ445716}, which tracks the distribution of Red Clump Giant stars~\citep{Nataf2013ApJ76988} that is obtained from surveys like OGLE/VVV~\citep{Minniti2010NewA15}:
\begin{equation}
 \rho_{\rm BB} = \rho_0 \cdot K_0(r_{\rm BB}) \, 
    \label{eq:BB}
\end{equation}
where $K_0(r_{\rm BB})$ is the Bessel function of second kind and order 0, and $$r_{\rm BB} =  \left( \left[ \left(\frac{x}{x_0}\right)^2 + \left(\frac{y}{y_0}\right)^2 \right]^2 + \left(\frac{z}{z_0}\right)^4\right)^{1/4},$$
where $x_0 = 67$~pc, $y_0 = 29$~pc, and $z_0 = 27$~pc.
This distribution is the most popular one for studies related to the Galactic center excess, and was found to agree well with the 511 keV signal distribution~\citep{Siegert2022MNRAS509L11}.

%\newline

For the nuclear bulge, we consider two components: 

\textbf{Nuclear stellar disk (NSD):}
\begin{equation}
\rho_{\rm NSD} (r) = \begin{cases}
     \rho_0\left(\frac{r}{1 \text{ pc}}\right)^{-0.1} \cdot e^{(-\frac{|z|}{45 \text{ pc}})} & \text{$r < 120$ pc} \\
     \rho_1\left(\frac{r}{1 \text{ pc}}\right)^{-3.5} \cdot e^{(-\frac{|z|}{45 \text{ pc}})} & \text{$120\text{ pc} < r < \text{220 pc}$} \\
     \rho_2\left(\frac{r}{1 \text{ pc}}\right)^{-10} \cdot e^{(-\frac{|z|}{45 \text{ pc}})}, & \text{$r > 220$ pc ,}
  \end{cases} 
\label{eq:rhoNSD} 
\end{equation}
where $r$ stands for the distance from the center at a given height ($z$ coordinate) (i.e. $r = \sqrt{x^2 + y^2}$). The normalizations at different $r$ are set such as the function is continuous at the breaks~\citep{Launhardt_2002}. 

\textbf{Nuclear stellar cluster (NSC):}
\begin{equation}
\rho_{\rm NSC} (r) = \begin{cases}
     \rho_0 \left(1 + \left(\frac{r}{r_0}\right)^2\right)^{-1} & \text{$r < 6$ pc} \\
     \rho_1\left(1 + \left(\frac{r}{r_0}\right)^3\right)^{-1}  & \text{$6\text{ pc} < r < \text{200 pc}$} \\
     0, & \text{$r > 200$ pc}
  \end{cases} 
\label{eq:rhoNSC} 
\end{equation}
where the normalizations at different $r$ are set such as the function is continuous at the breaks. Here, $r$ represents the spherically symmetric distance from the center of the Galaxy ($r = \sqrt{x^2 + y^2 + z^2}$) and $r_0 = 0.22$~pc~\citep{Portail2015L66}.

In what follows, we will either use these bulge components separately or combine the boxy-bulge distribution with the nuclear bulge.
The latter combination is indeed a more complete description of the source distribution at the bulge. In this case, even though we still normalize the total positron injection rate to the one corresponding to the $511$~keV observations as we will explain below, the relative normalization of each component will be fixed by the stellar distributions. To do so, we follow~\cite{Bartels_2018} and set the normalizations of the different components to be:
For the boxy-bulge, $\rho_0^{\text{BB}}=52.1$ M$_\odot$ pc$^{-3}$ is set such as the total bulge mass is $9.1\cdot10^{9}$~M$_\odot$~\citep{Licquia_2015}. $\rho_{0}^{\text{NSD}} = 301$ M$_\odot$ pc$^{-3}$ to get a total mass enclosed in the inner $120$~pc of $8\cdot10^8$~M$_\odot $~\citep{Launhardt_2002}. $\rho_0^{\text{NSC}}= 3.27\times10^{6}$~M$_{\odot}$/pc$^{3}$, such that the total mass enclosed in the inner $625$~pc is $7\cdot10^5$~M$_\odot$~\citep{Launhardt_2002}.

In Figure~\ref{fig:Distribs}, we show a comparison of the different distributions discussed above in logarithmic scale to show the innermost region around the GC. The normalization of each component is the one commented above. In this panel, we indicate with two different lines the density trend with distance from the GC in the two perpendicular directions at the Galactic plane (X and Y, in the labels), given that the boxy-bulge template is non radially symmetrical. A similar figure, but including also a Navarro-Frenk-White-like dark matter density distribution is shown in Appendix~\ref{sec:App}, Figure~\ref{fig:Distribs_wDM}.

\subsection{Positron injection spectra}
\label{sec:PosInj}

As discussed above, the injection of positrons from different sources might have different spectral trends with energy, from line-like injections to power-law trends. 
We here consider three different spectral behaviors: a first, phenomenological one, a second one corresponding to more realistic radionuclides spectra, and a third one 
representing pulsar-like spectra.

\begin{figure}
    \includegraphics[width=\columnwidth]{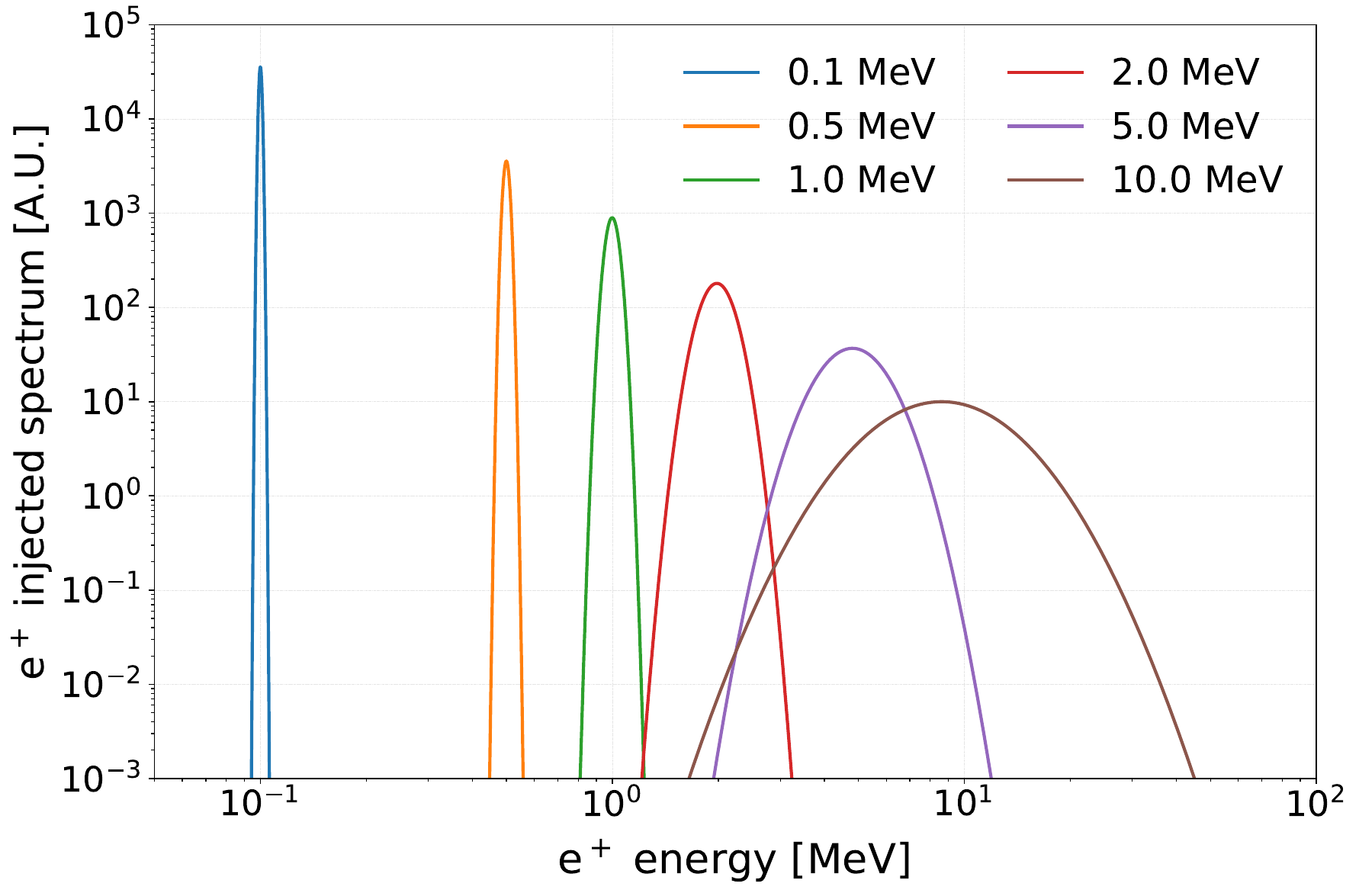}
    \includegraphics[width=\columnwidth]{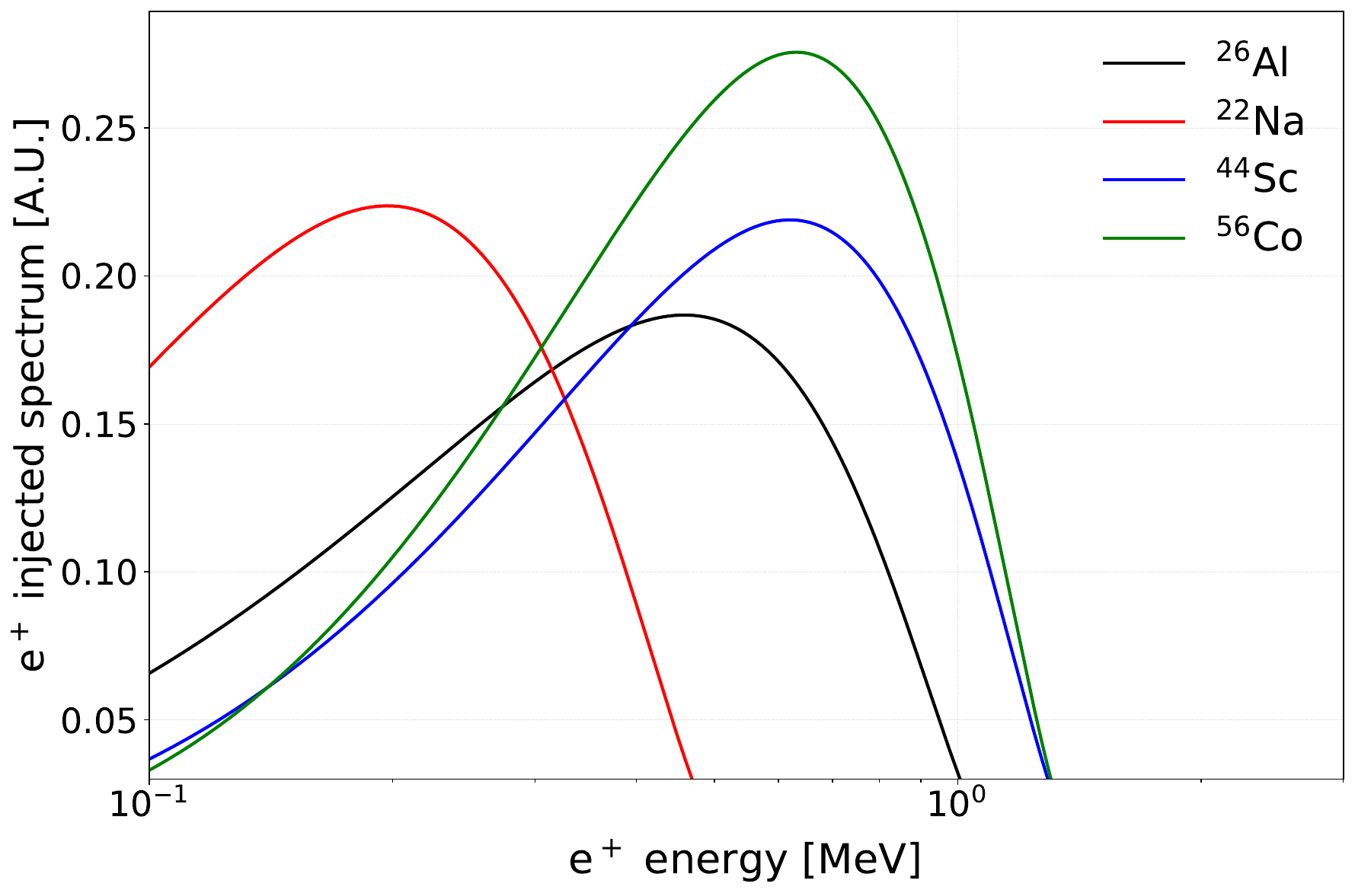}
    \caption{\textbf{Top panel:} Positron injection spectra of the signals described by Eq.~\ref{eq:LogInj} for different mean energies (indicated by the colors in the legend). The width of each spectrum is fixed to one-fiftieth of its mean energy. \textbf{Bottom panel:} Injection spectra for different $\beta^+$ radionuclides commonly associated to massive stars, SNe and novae, as indicated in the legend.}
    \label{fig:Injection}
\end{figure}

\textbf{Log-normal spectrum.}
We first test a general case where the positron injection is modeled following a log-normal spectrum (upper panel of Fig.~\ref{fig:Injection}). This has 
the advantage to be flexible enough to capture the behavior expected from injection of radionuclides (see lower panel of Fig.~\ref{fig:Injection}) and also line-like injection features like the ones expected from dark matter. 
The main parameter (mostly affecting the ionization phenomenology) is here the mean energy of the injected positrons. 
%We find that this type of injection spectrum leads to an ionization rate that depends primarily on the mean energy of the injected positrons (as well as on the spatial distribution of the signal, as discussed below). This behavior arises from the energy dependence of positron energy losses, which ultimately determines the spectral distribution of positrons capable of ionizing the gas in the CMZ.
%. The reason to implement this general injection is because we found that the ionization rate, in this case, mostly depends on the injected peak positron energy (and the spatial distribution of the signal, as we show below), which is due to the dependence of the energy losses on energy, that ultimately dictates the spectral distribution of positrons that are able to ionize the gas in the CMZ.
%We show later (Sec.~\ref{sec:Results}) the predicted ionization rates from other spectral trends, which do not vary significantly.
We parameterize this log-normal injection as:
\begin{center}
\begin{minipage}{0.85\linewidth}
\begin{equation}
  \frac{dN_{e^+}}{dE} \propto \frac{1}{E\,\sigma_E\,\sqrt{2\pi}} \,\,\text{exp}\left[ - \left(   \frac{\ln(E) - \ln(E_{\text{mean}}) }{2\sigma_E }\right)^2 \right]
   \,,
\label{eq:LogInj}
\end{equation}
\end{minipage}
\end{center}
$\sigma_E$ is given by imposing the width of the spectrum to be a fraction of the mean energy, i.e. standard deviation is E$_\text{mean}\times f$, where $f$ is the fraction of the mean energy that sets how wide is the signal. From this, $\sigma_E$ is defined as $\sigma_E =  \sqrt{\ln(1 + 
(\text{E}_\text{mean}\times f)^2)}$~\citep{Aitchison1957, Limpert2001}. % where the fraction is the number that we want to use... %%\sqrt{log( 1 + \times f)^2 ))}$

Here, we set standard deviation of the log-normal distribution to be 1/50 of the mean energy (i.e. $f = 1/50$ is used as default). We have tested that changing this value from 1/500 to 1 times the mean energy leads to changes that are not higher than $\sim15\%$ in the total ionization rate predicted. The reason for this, is because the signal is always normalized to the total rate of positrons injected in the inner $8$~deg, as we describe in Sect.~\ref{sec:Normaliz}.
In these calculations, we have considered values of E${_\text{mean}}$ from $0.1$ MeV to $20$~MeV for the reasons that will be discussed later. 
In the example shown in Fig.~\ref{fig:Injection}, we represent, in the upper panel, the general signals (following a log-normal distribution) for different positron mean energies. For concreteness, the $x$-axis represents the positron kinetic energy.

%For comparison, we also show the spectrum expected from different $\beta^+$ radionuclides in the lower panel: $^{26}$Al representative from injections from massive stars and with a endpoint kinetic energy of $1.7$~MeV, $^{22}$Na commonly produced in novae and with an end-point energy around $0.5$~MeV $^{44}$Sc produced from the decay of $^{44}$Ti) in core-collapse SNe, and $^{56}$Co which is produced in SNe-Ia (the two latter have similar end-point energies of $E_{e^+}^{\rm max}\sim1.2$~MeV). Their injection spectrum is described around Eq.~\ref{eq:beta}~\citep{ContiEriksson2016, MIRD1979, LandstreetBetaGammaNotes}.
%We have also tested the case of a power-law injection, with cutoff at around a few MeV, to be consistent with the COMPTEL signal found in~\cite{Kn_dlseder_2025}, which requires positrons injected with a maximum energy at around a few MeV (see Sect.~\ref{sec:RadPL} for details). As mentioned above, this injection is characteristic of pulsars, X-ray binaries or microquasars.

\textbf{$\beta^+$-radionuclides spectrum.}
The spectrum of $\beta^+$ unstable isotopes produced in stars, SNe or novae, such as $^{56}$Co, $^{44}$Ti, $^{26}$Al, or $^{22}$Na, can be approximated as~\citep{ContiEriksson2016, MIRD1979, LandstreetBetaGammaNotes}:
\begin{center}
\begin{minipage}{0.85\linewidth}
\begin{equation}
    \frac{dN_{e^+}}{dE_{e^+}}
    \propto \sqrt{E_{e^+}^2 + 2E_{e^+}m_{e}} \cdot (Q-E_{e^+})^2 \cdot (E_{e^+} + m_e) \cdot F(E_{e^+}, Z) \,\,\, .
\label{eq:beta}
\end{equation}
\end{minipage}
\end{center}
Note that the proportionality constant is not relevant here, since we will require the integrated spectrum to yield the required positron injections rate to explain the 511 keV signal.
$Q$ represents the energy yield of the transition and as such is the upper bound on the kinetic energy of the positron, E$_{e^+}$, and $Z$ is the atomic number of the radionuclide. For the cases that we study, we use Q$_{^{26}\rm Al} = 1.173$~MeV, Q$_{^{22}\rm Na} = 0.5455$~MeV, Q$_{^{44}\rm Sc} = 1.474$~MeV and Q$_{^{56}\rm Co} = 1.46$~MeV.
The Fermi function, F$(E_{e^+}, Z)$ is approximately given by~\citep{LandstreetBetaGammaNotes}
\begin{center}
\begin{minipage}{0.85\linewidth}
\begin{equation}
   F\left(E_{e^+}, Z \right)
    \simeq \frac{2\pi\eta}{1 - \text{exp}(-2\pi\eta)} \,\,\, ,
\label{eq:Fermi}
\end{equation}
\end{minipage}
\end{center}
with $\eta = -Z\alpha_c/c\beta_e$, where, $\alpha_c$ is the fine structure constant and $c\beta_e$ is the positron speed in speed of light units.
We show the spectra for each of these nuclei in the bottom-left of Fig.~\ref{fig:Injection} in arbitrary units, and also in Fig.~\ref{fig:InjSpectra_2} but with the normalization of the injection spectrum that leads to the adopted positron rate within the bulge. 
Compared with the phenomenological spectrum, these injections are similar to the log-normal parametrization for positrons with kinetic energy $\sim1$~MeV and a large $\sigma_E$ value (leading to broader signals). 

\textbf{Pulsar spectrum.}
Finally, we explore the case of injection of low energy electrons and positrons injected following power-law spectra, as expected from pulsars or X-ray binaries~\citep{Guessoum_2006, bandyopadhyay2009origin}. 
Since we normalize our spectrum to the total injected positron rate (see below), the power-law requires setting a minimum positron energy (otherwise all electrons and positrons are injected at kinetic energies below the minimum ionization energy, that is about $15$~eV).  Therefore, we parametrize this injection as
\begin{center}
\begin{minipage}{0.85\linewidth}
\begin{equation}
    \frac{dN_{e^+}}{dE_{e^+}}
    \propto \left(\frac{E}{E_0}\right)^{-\gamma} \,\,\text{exp}\left(-E/E_{\rm cut}\right) \,\,\, ,
\label{eq:PL}
\end{equation}
\end{minipage}
\end{center}
where E$_0$ is set to $0.1$~MeV and varied to see its impact. 
We show this injection in the lower-left panel of Fig.~\ref{fig:InjSpectra_2}, where the legend represents different values of the cutoff energy and we use $\gamma=2.2$ there.

\subsection{511 keV line signal normalization}
\label{sec:Normaliz}
After injection, positrons interact with the gas and propagate, eventually becoming thermal and finding ambient electrons with which they interact and produce $511$~keV photons. Given the low energy of these positrons %and the high density of the CMZ (n$^{\text{CMZ}}_{H_2}\sim150$~cm$^{-3}$)
the dominant time-scale for these positrons is that of energy losses from ionization of the gas, such that diffusion can be neglected (see e.g. Fig.~$3$ in the Appendix of~\cite{DelaTorreLuque:2024fcc} for the comparison between different time scales in the GC). We note that this approximation is valid for low energy particles -- below a few tens of MeV -- and in dense media, and it has the advantage that it does not need to assume a diffusion coefficient or other propagation parameters that are not known at all near the GC. Including diffusion and/or advection would slightly flatten the positron spatial distribution compared to what is shown in Fig.~\ref{fig:Distribs}. 
Under this assumption, the steady-state distribution of positrons can be calculated as 
\begin{center}
\begin{minipage}{0.85\linewidth}
\begin{equation}
    \frac{dn_{e^+}}{dE}\left(E, \mathbf{x} \right)
    = \frac{1}{b_e(E, n_{H_2})}
      \int Q_e(E', \mathbf{x})\, dE' ,
\label{eq:fluxe}
\end{equation}
\end{minipage}
\end{center}
with $b_e$ as the energy-loss term from ionization ($b_e \equiv (dE/dt)_\textrm{ion}$), calculated following the one implemented in the {\tt DRAGON2} code (see Eq.~(C.39) of~\cite{DRAGON2-1}), $n_{H_2}^\textrm{CMZ}=150$~cm$^{-3}$ is taken as the mean density of the CMZ~\citep{Ravikularaman2025AandA694A114, Indriolo_2015}, using a mixture of $H$ and $He$ of 1/10, and $Q_e\left(E_e, \textbf{x} \right) = Q_0\cdot \rho_{\text{Bulge}}(x, y, z) \cdot \frac{dN}{dE_e}(E_e)$ is the source term (injection spectrum). 
Here, the normalization, Q$_0$ is described in the next section, $\frac{dN}{dE_e}(E_e)$ is the positron injection and $\rho_{\rm Bulge}$ is the density distribution described above (which comes from the sum of the different components). Note that this source term follows the distribution of sources in the bulge and, thus, it makes the spatial distribution of positrons to follow the bulge density distribution.

Variations in the gas density distribution could lead to variations in the distribution of electrons and positrons and, hence, in the ionization rate profile. Our estimations assume a spatially uniform gas density throughout the CMZ, which is a reasonable approximation on scales beyond the circumnuclear disk (i.e. above 1-5 pc). Nevertheless, observations reveal a marked longitudinal asymmetry, characterized by higher densities at positive longitudes, although no clear global trend in the gas density is evident. In turn, latitudinal asymmetries are relatively mild~\citep{Battersby2025_CMZ_overview}. Moreover, dense clumps and cores are present throughout this region, their volume filling factor is small and they do not significantly affect this assumption.

\begin{figure*}
%\textbf{BoxyBulge}
\includegraphics[width=\columnwidth]{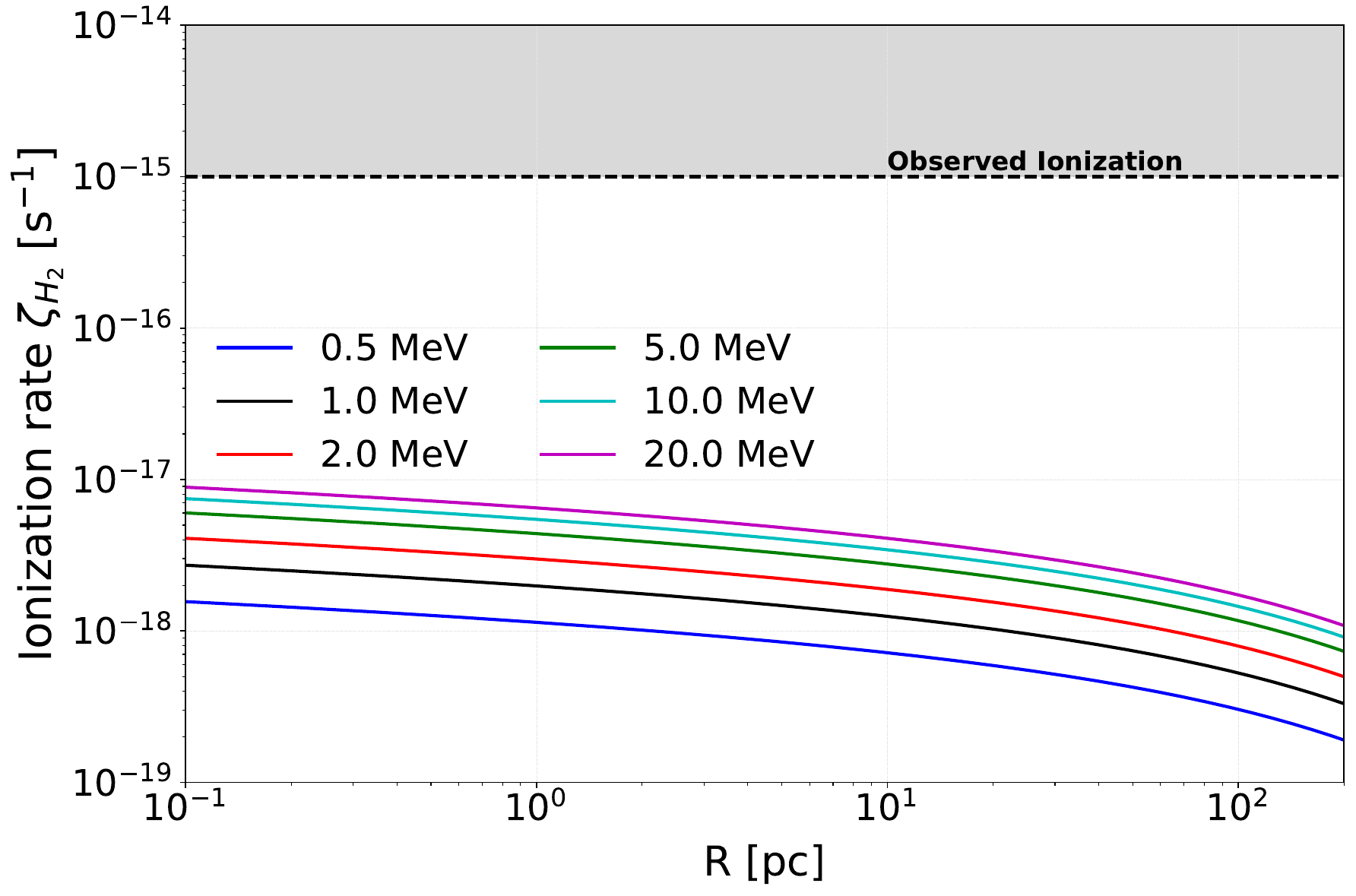}
%\textbf{NSD}
\includegraphics[width=\columnwidth]{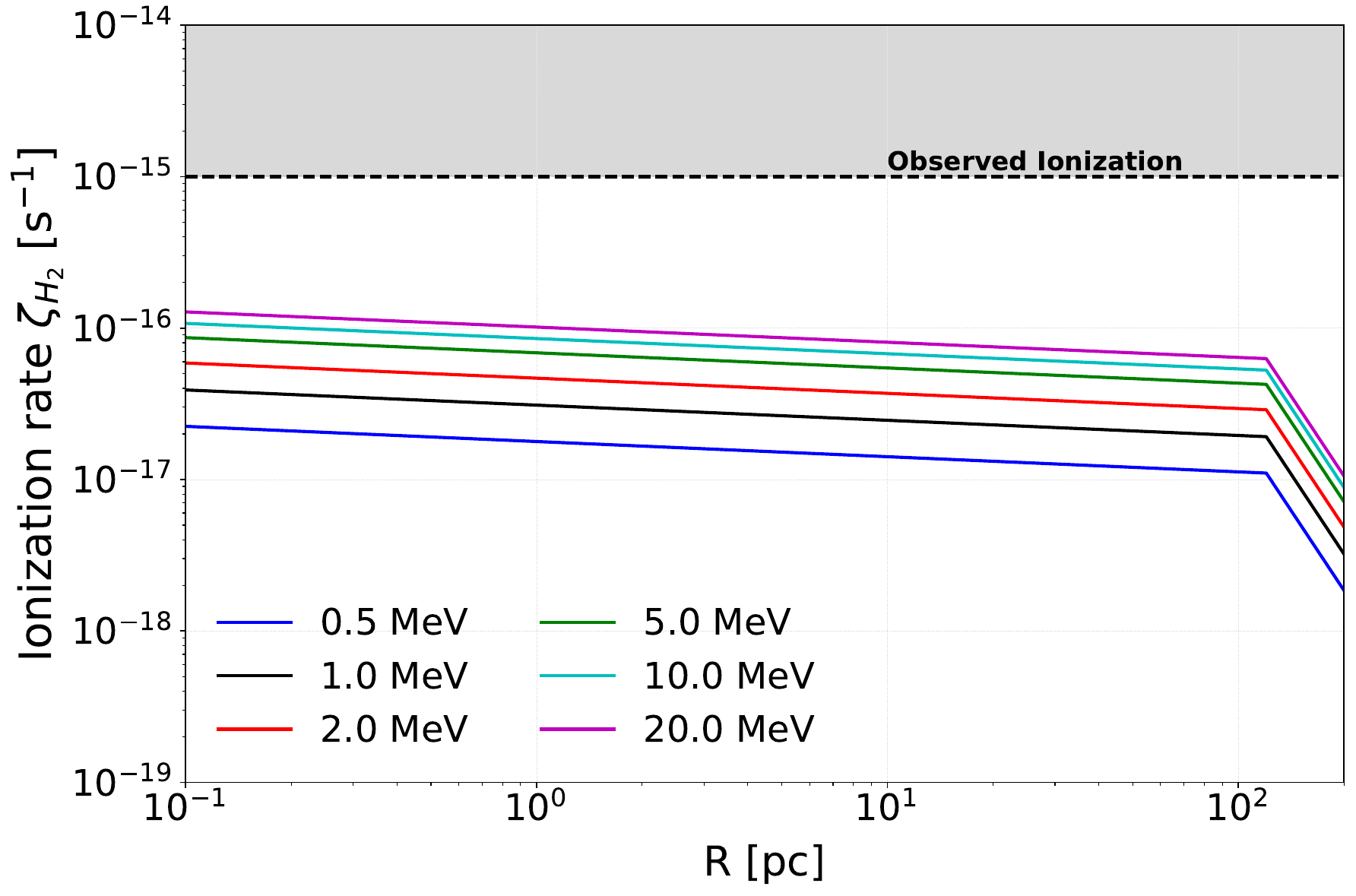}
%\textbf{BoxyBulge+NSD}
\includegraphics[width=\columnwidth]{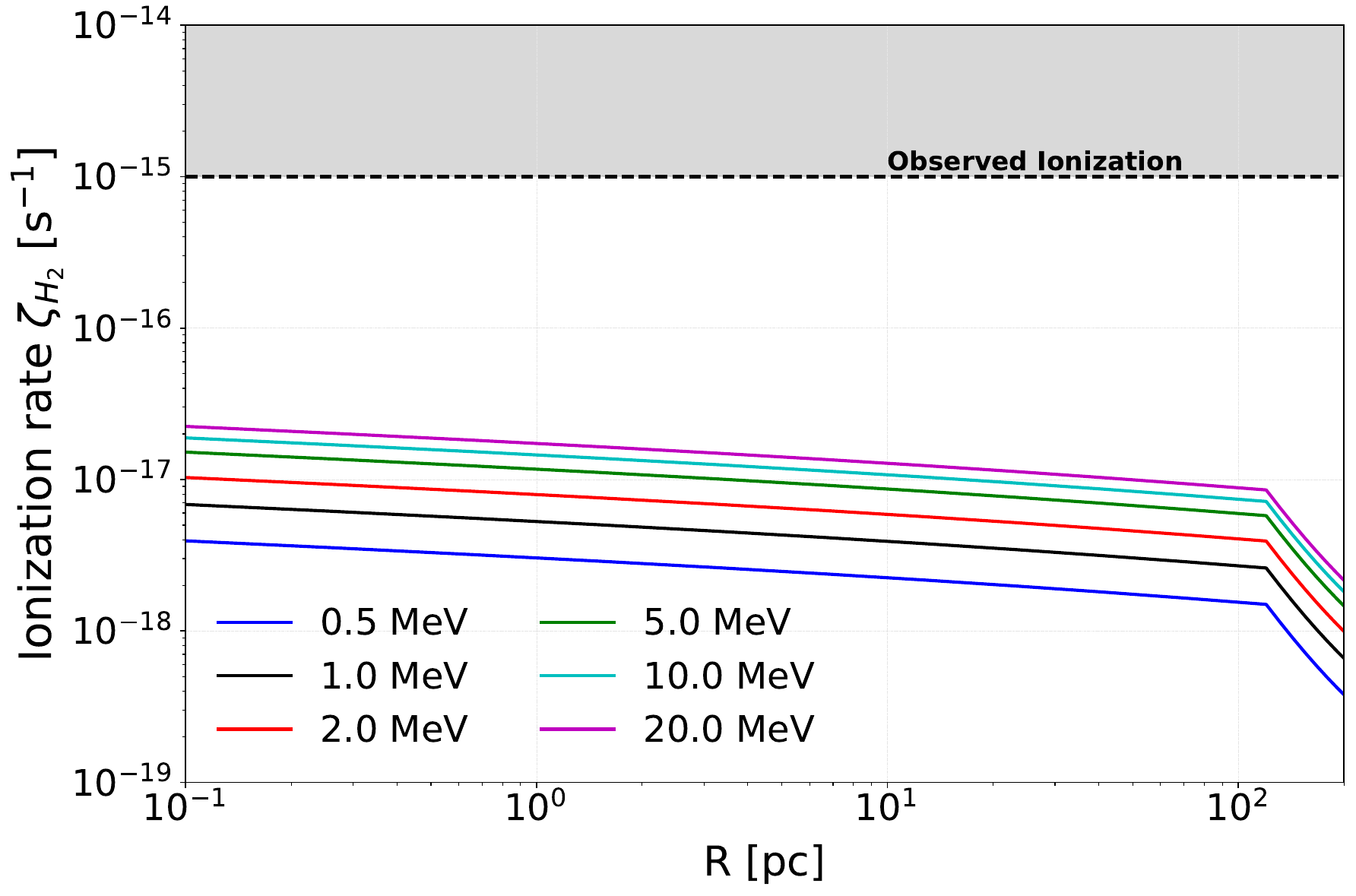}
%\textbf{BoxyBulge+NSD+NSC}
\includegraphics[width=\columnwidth]{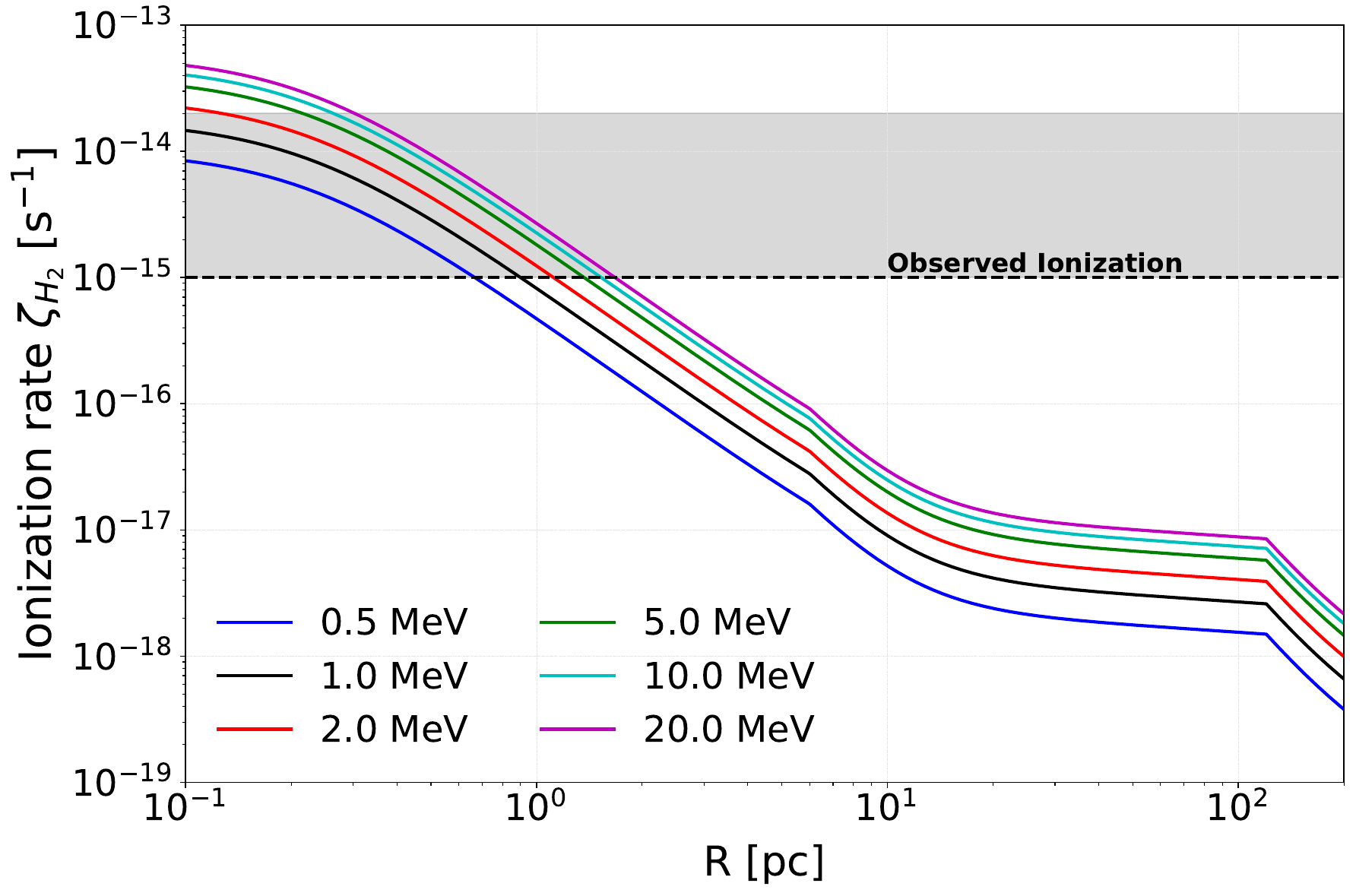}
    \caption{H$_2$ ionization rate expected at the CMZ from positron sources following different source density distributions and assuming an log-normal injection spectrum. \textbf{Top-left panel:} H$_2$ ionization rate expected from positron sources following the boxy-bulge distribution. \textbf{Top-right:} Ionization rate expected from positron sources following the nuclear stellar disk (NSD). \textbf{Bottom-left:} Expected ionization rate in the CMZ from a combination of the nuclear stellar disk and boxy-bulge distributions. \textbf{Bottom-right:} Ionization rate at the CMZ from a combination of the nuclear stellar disk, nuclear stellar cluster and the boxy-bulge distributions.}
    \label{fig:Ionizations}
\end{figure*}

In order to link the expected CMZ ionization rate with the $511$~keV observations, we normalize the positron spectra obtained by imposing the rate of positrons required to explain the $511$~keV observations of the bulge. Therefore, we impose that the positron injection rate from these sources\footnote{Given that observations support a constant flux of 511 keV photons (in time), the injection rate must match the rate of positron annihilation.} is $2\cdot 10^{43}$~e$^+$/s in the inner $8$~deg~\citep{Siegert_2016, Skinner2015Integral2014_511keV}. Specifically, we impose this positron injection rate within the volume enclosed in the inner $1.15$~kpc of the GC, which corresponds to an angular radius of $8$~deg in the inner Galaxy. 
Said in other words, Q$_0$ is obtained by integrating the positron flux over energy and over the inner $8$~deg (the bulge) to provide a positron rate of $2\cdot 10^{43}$~e$^+$/s.

\begin{center}
\begin{minipage}{0.85\linewidth}
\begin{equation}
\begin{split}
    \int_{\rm Bulge} dV \int dE \, Q \left(E, \mathbf{x} \right) = \\
&\hspace{-3cm} Q_0\int_{\rm Bulge} dV \rho_{\text{\rm Bulge}} \int dE \,  \frac{dN}{dE}\left(E, \mathbf{x} \right)
    = 2\cdot 10^{43} \, \text{e$^+$/s} ,
    \end{split}
    \label{eq:Norm}
\end{equation}
\end{minipage}
\end{center}

\begin{comment}
\begin{center}
\begin{minipage}{0.85\linewidth}
\begin{equation}
\begin{split}
    \int_{\rm Bulge} dV \int dE \, \frac{dn_{e^+}}{dEdt} \left(E, \mathbf{x} \right) =  \\ 
&\hspace{-3cm} \int_{\rm Bulge} dV \int dE \, \frac{dn_{e^+}}{dE}\cdot \frac{1}{\tau_{Ion}}
    = 2\cdot 10^{43} \, \text{e$^+$/s} ,
    \end{split}
    \label{eq:Norm}
\end{equation}
\end{minipage}
\end{center}
where $\tau_{Ion}$ is the dominant time-scale for MeV electrons, which is the ionization energy-loss time-scale.
\end{comment}
This normalization ensures that the total ionization rate would not change even including propagation effect (provided that the positron rate to which we normalize is the total number of positrons in the bulge), although the ionization profile could slightly flatten, as commented above.

\subsection{Ionization rate in the CMZ}
\label{sec:ioniz}
Given the positron density within the CMZ, calculated from Eq.~\ref{eq:fluxe}, the positron flux, $J(E,\textbf{x})$, can be computed as
\begin{center}
\begin{minipage}{0.85\linewidth}
\begin{equation}
    J_e(E, \textbf{x}) = \frac{dn_{e^+}}{dE}\left(E, \textbf{x} \right) \frac{\beta_e(E)c}{4\pi}
     \,,
\label{eq:flux}
\end{equation}
\end{minipage}
\end{center}
where $\beta_e(E) c$ is the speed of the positron. 
Finally, with this flux we compute the rate of ionization of molecular hydrogen within the CMZ as~\citep{Padovani_2009, Phan_2023}
\begin{center}
\begin{minipage}{0.85\linewidth}
\begin{equation}
  \zeta (\textbf{x}) = 4\pi \int^{E_\textrm{max}}_{E_\textrm{min}} J_e(E, \textbf{x}) \, \sigma(E)(1 + \theta_e(E)) dE 
   \,,
\label{eq:main}
\end{equation}
\end{minipage}
\end{center}
where $E_\textrm{max}$ is the maximum kinetic energy with which the positrons are injected and $E_\textrm{min}$ is the minimum kinetic energy to produce ionization in the gas ($15.43$~eV~\cite{Padovani_2009, Phan_2023}), $\sigma(E)$ is the $e$-H$_2$ cross-section, calculated as in Eq.~(7) of~\cite{Padovani_2009} and $\theta_e(E)$ is the number of secondary ionizations per primary ionization, calculated as Eq.~(2.23) of~\cite{krause2015crimecosmicray}. Note that the spatial dependence of the ionization rate is inherited by the distribution of the injected electrons and positrons.

The calculations above can be applied in general for any spectral dependence of the source term.

\section{Results}
\label{sec:Results}
In this section, we present our results, derived for different components of the bulge and their combination, as well as for different injection spectra based on the potential sources of the $511$~keV excess discussed above.

\subsection{General case: log-normal injection spectrum of positrons}

Following the calculation described above, we show, first, in Fig.~\ref{fig:Ionizations} the expected ionization rate in the inner few hundreds parsecs of the Galaxy (corresponding to the CMZ region) for the general case of log-normal injection spectrum of positrons, following the different source density distributions commented in Sect.~\ref{sec:BulgeDist}. The different colors represent the ionization rate expected for positrons injected with the mean energy indicated in the legend in every case.
The top-left panel refers to the ionization rate calculated assuming only the boxy-bulge distribution (Eq.~\ref{eq:BB}), which is the dominant one at large distances at the edges of the CMZ. 
The resulting ionization rates are very similar to those obtained when assuming a symmetric bulge distributed of~\cite{Korol_2018}, since the two models provide very similar descriptions of the bulge. In the upper-right panel, we display the radial profile of the ionization rate calculated using the NSD source-density distribution (Eq.~\ref{eq:rhoNSD}). This distribution constitutes the main mass component within the inner $200$~pc.

As we see, in both cases the predicted level of ionization rate is significantly below $\sim10^{-15}$~s$^{-1}$, which is around the minimum value required to explain the abundances of ionized hydrogen~\citep{Oka_2005, Oka_2020, Geballe_2010, Indriolo_2009}.
However, although the expected ionization rate is much lower than observed, the values of ionization rate obtained adopting the NSD distribution are significantly larger than those from the boxy-bulge case,  which is due to the fact that the source density in the inner Galaxy is much larger than the density outside the CMZ for the NSD distribution. Since the total positron injection rate is normalized to the one required to reproduce the $511$~keV intensity in the whole bulge (Eq.~\ref{eq:Norm}), this makes most of the injected positrons be accumulated in the inner $\sim200$~pc. 
In other words, the larger the density in the inner region of the bulge over the outer region, the larger is the expected ionization rate in the CMZ. 

In these panels, one can see that, for a fixed rate of injected positrons, the higher is the energy of the positrons injected, the more they can ionize the medium, in general. This is because these positrons deposit all their energy within the CMZ and they become thermal before escaping this region. As we discussed above, this is true only for low energy positrons injected, below a few tens of MeV, where the diffusion time-scale is subdominant, and it is the reason because we restrict ourselves to a maximum positron injection energy of $20$~MeV. 
Moreover, we checked the effect of adopting different values for the variance of the log-normal distributions describing the positrons injection spectrum, and found that the ionization rate for each mean positron energy, remains very similar (with variations of the order of $15\%$, increasing slightly toward higher values of the variance). This is because the integrated signal is normalized to the total positron positron rate, so that making it broader or tighter does not change the total positron flux, only its spectral shape (which is why there are small variations).

\begin{figure*}

\includegraphics[width=\columnwidth]{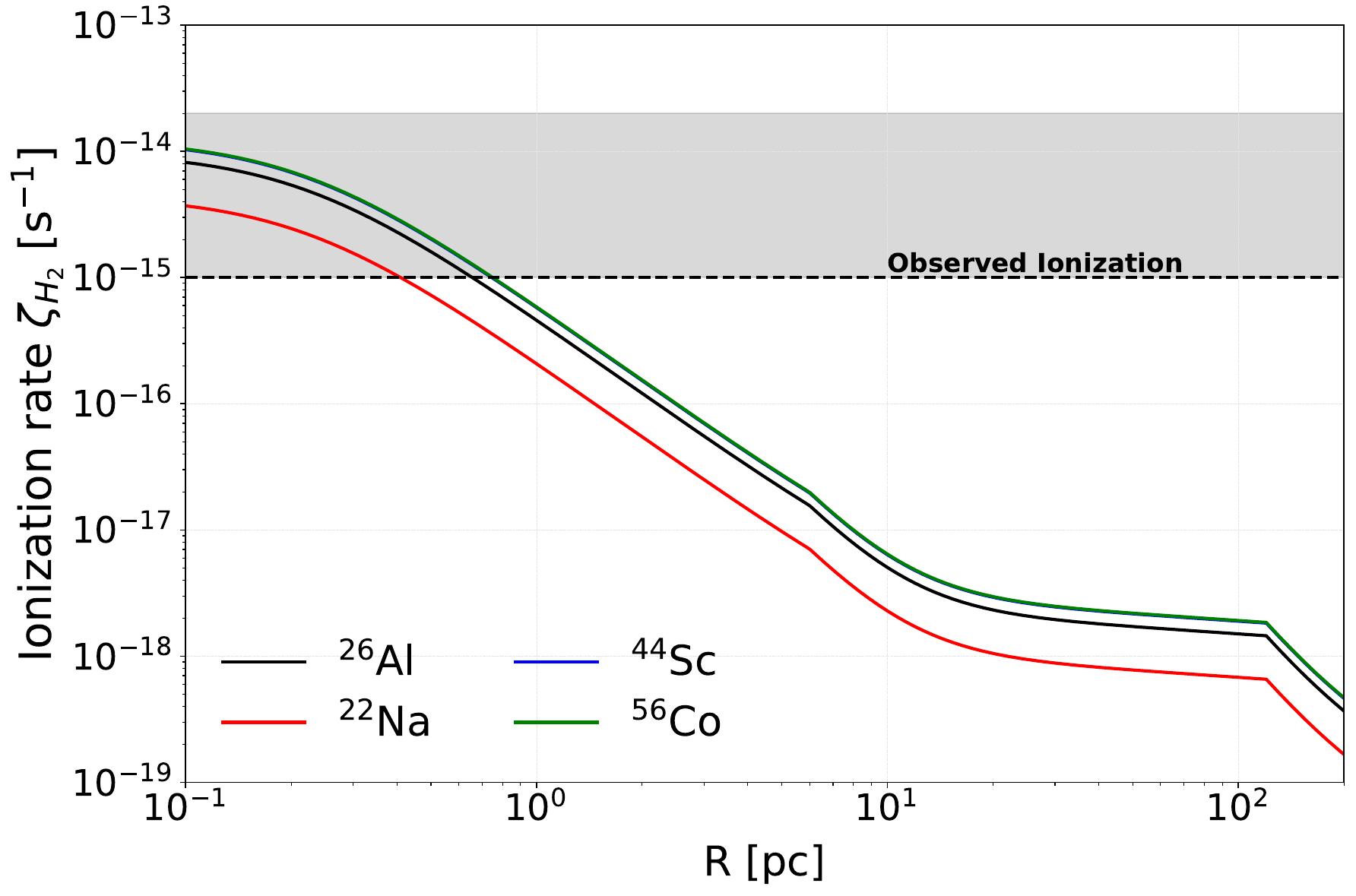}
\includegraphics[width=\columnwidth]{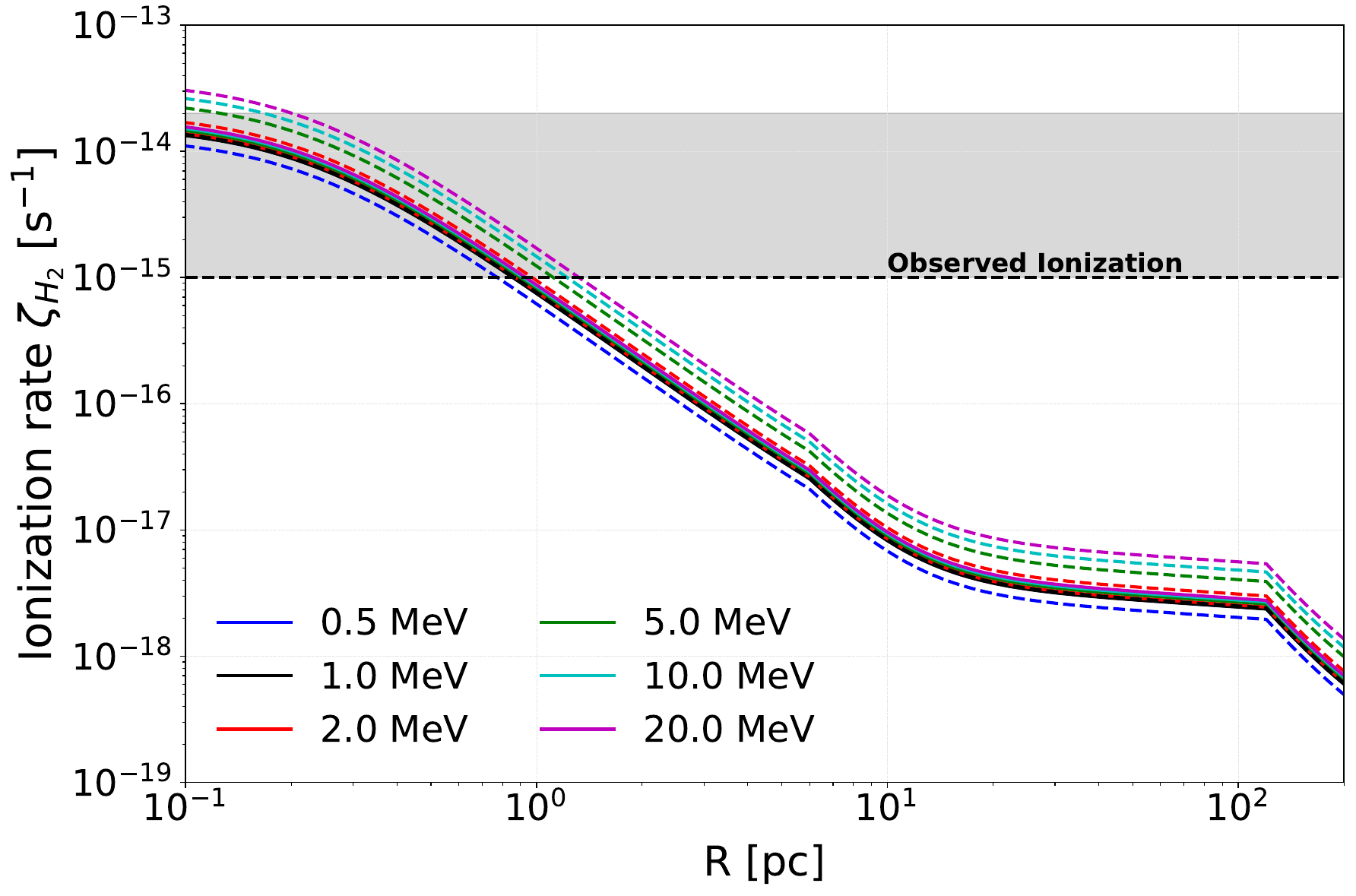}
    \caption{H$_2$ ionization rate expected at the CMZ from positron sources following a combination of the nuclear stellar disk, nuclear stellar cluster and the boxy-bulge distributions and assuming radionuclides injecting the positrons (left panel) and a power-law positron injection (right panel). In the left panel, the different colors indicate different radionuclides. In the right panel, the different colors indicate different cut-off energies. In this panel solid lines correspond to a power-law where the spectral index is $\gamma = 2.2$, while the dashed lines correspond to a power-law of $\gamma = 1.2$.}
    \label{fig:Ionizations_2}
\end{figure*}

Then, the bottom-left of Fig.~\ref{fig:Ionizations} depicts the profile of the ionization rate for a combination of NSD and boxy-bulge distribution, while the bottom-right the combination of  boxy-bulge with NSD and NSC distributions, which is expected to be the most realistic case. In both cases, we normalize the total injection rate after adopting the normalizations ($\rho_0$ values) discussed above, which sets the relative importance of each component (i.e. the total density of the summed components is not relevant but their relative contributions are very important for the signal's morphology).
As we see, the inclusion of the NSC changes significantly the expected ionization rate. Without this component, the ionization profile is expected to be very uniform in all cases, with variations of less than a factor of a few. In contrast, the inclusion of the NSC significantly boosts the ionization rate in the inner parsec, bringing it to -- or even above -- the observed level.
However, this also causes the expected ionization profile to vary strongly with distance from the GC, changing very quickly. Although radial winds and positron diffusion would tend to flatten the profile, the expected ionization varies by 2–3 orders of magnitude beyond the inner parsec\footnote{the resolution for measurements of rovibrational lines of ionized molecules -- in particular, in the infrared and for H$^+_3$ -- are typically above a fraction of a parsec}, making the profile too steep to match the observations. This conclusion is further supported from what we show in Sect.~\ref{sec:RadPL}.
A detailed study of the effects of such winds would be necessary to reach a definitive conclusion.
Remarkably, we note that these values are much higher than those usually expected from cosmic rays penetrating the CMZ, or even than other sources of high-energy particles known to be in this region~\citep{Ravikularaman2025AandA694A114}.

\subsection{Specific cases for the positron injection spectrum}
\label{sec:RadPL}

In this section, we adopt specific spectral distributions for the injected positrons, based on two general parametrizations of the potential $511$~keV sources described above: The general spectrum of $\beta^+$ decay from radionuclides and a power-law with an exponential cutoff.

As we see from the left panel of Fig.~\ref{fig:Ionizations_2}, positrons from radionuclides would lead to a similar level of ionization to the one presented in the generic spectrum shown above for a mean injection energy of around $1$~MeV, which is clearly insufficient to explain the ionization rate level observed in the CMZ.
However, we stress that in the processes in which these radionuclides are expected to be produced, high-energy particles—such as protons and heavier nuclei—are generated as well. These particles should also contribute significantly to the total ionization, and they are expected to follow a similar ionization profile. Therefore, although positron injection by itself cannot explain the observed level of ionization, this does not imply that the sources producing these positrons are incapable of accounting for the measured ionization.

The expected ionization rate profile from pulsar-like injection spectra is illustrated in the right panel of Fig.~\ref{fig:Ionizations_2}, where different colors indicate different cutoff energies. In this case, we assume that pulsars-like sources inject same numbers of positrons and electrons, and therefore we take into account an extra factor of 2 in the calculation of the ionization rate. Here, we find that the spectral index slightly affects the predicted ionization rate (see the difference between dashed and solid lines, representing the  case of $\gamma=1.2$ and  $\gamma=2.2$, respectively). Furthermore, the impact of the cutoff energy depends on the spectral index used:  
A larger $\gamma$ value (in absolute value) means that more positrons are distributed at the lower energies, which tends to make the estimation rate mainly driven by the value of E$_0$ that we use. Meanwhile, a low spectral index value (i.e. a harder spectrum) distributes more positrons at the higher energies, making the ionization rate more dependent on the cutoff energy used.
Adopting a E$_0$ value of $1$~MeV, instead of $0.1$~MeV, leads to an increase of around a factor of $\sim2$. In sum, all these variations predict an ionization rate similar to the one that we see in the previous cases, where the inclusion of the NSC increases significantly the ionization level in the inner parsec, leading to a very steep profile. 

As a last note here, we stress that although pulsars are likely to explain the high-energy part of the local positron flux measured by PAMELA and AMS-02~\citep{Aguilar2014AMS02Positrons}, simulations indicate that positrons from pulsars and millisecond pulsars cannot efficiently escape and be injected into the ISM following a power-law with energies below $\sim$a few tens of MeV (see e.g.~\cite{2010MNRAS.408.2092T}), making them less suitable to explain the low energy ($\sim$MeV injection) required to explain the COMPTEL signal commented above. On top of this, observations of TeV halos~\citep{Abeysekara2017ApJ84340} support a very high-energy injection of electrons and positrons from pulsars, which would propagate long distances before thermalizing (with path-lengths to become thermal much larger than for MeV positrons~\citep{Siegert2022MNRAS509L11, laTorreLuquePedro:2024est}), making them not suitable to explain the concentration of the 511~keV excess around the GC. Additionally, such high-energy injection would violate constraints from in-flight annihilation and diffuse emission~\citep{Sizun2006PRD74_063514, Beacom_2006, Das2025arXiv2506.00847}. Meanwhile, X-ray binaries could inject more efficiently MeV positrons and even contribute significantly to the sub-MeV continuum diffuse emission, as shown in~\cite{Kantzas_2024}.

\section{Discussion and conclusions}
\label{sec:Conclusions}
In this paper, we investigated whether the positron injection rate required to explain the $511$~keV emission observed in the Galactic bulge can also explain, or significantly contribute to, the anomalous ionization rate measured in the CMZ. To this end, we considered several spatial distributions for potential sources of the 511 keV excess, motivated by the observed stellar distribution in the Galactic bulge, and explore different parameterizations of their positron injection spectra. Given that the observed morphology of the $511$~keV emission is very concentrated around the GC, and the recent hint for in-flight annihilation signal from MeV positrons detected in COMPTEL data, we focused on positrons injected below a few tens of MeV.

We found that, if the positron emission follows the boxy bulge, the NSD or the combination of these two, the total level of ionization in the CMZ remains below what is observed, with maximum values for the ionization rate that are in the range of $10^{-16}$~s$^{-1}$ to $10^{-15}$~s$^{-1}$, but generally predicting a level of ionization that is around $10^{-17}$~s$^{-1}$ beyond the inner pc. 
A notable aspect of these results is that such source distributions yield a fairly uniform spatial ionization profile within the CMZ, matching one of the most striking features observed in the CMZ ionization rate. 
However, including the NSC, which is highly concentrated within the $\sim$inner parsec, the expected ionization level increases sharply in the inner parsec, reaching ionization rates even above $10^{-14}~\mathrm{s^{-1}}$, but then decreases rapidly at larger radii, falling significantly below the observed ionization level and producing a profile that is not compatible with the uniformity of CMZ measurements.
This holds true either for the case of radionuclides spectra or a power-law-like positron injection, and even for line-like injections, as long as the sources injecting these positrons follow the distributions of stars in the Galactic bulge. 
Additionally, we note that when combining the different profiles, we assume that the positron injection is directly proportional to the source density of each component, even though different components may host different source populations. Each of these components may hold different sources, that could change the spatial ionization profile. In this regard, the BB profile is expected to contain the older systems, while the NSC is characterized for having a higher star formation activity~\citep{Nie_2026}.  %, which could slightly modify the results. Nevertheless, this effect would only tend to make the ionization profile more closely follow the distribution of the dominant component.

%The case of dark matter injecting electrons and positrons was first explored in~\cite{DelaTorreLuque:2024fcc}, where the CMZ ionization from the direct annihilation of dark matter into $e^+e^-$ pairs, for different dark matter profiles was explored. Meanwhile,~\cite{balaji2025exciteddarkmattersolution}, whose conclusions are similar to the ones that we find here, carried out a combined analysis of the $511$~keV emission and the CMZ ionization rate, within a more specific dark matter scenario, referred to as Excited dark matter~\citep{FinkbeinerWeiner2007, Cappiello_2023}: in this scenario, collisions between dark matter particles inelastically excite one of the dark matter particles to a slightly heavier state that decays back producing an $e^+e^-$ pair. This study shows that the expected ionization rate from dark matter would fall short to explain the anomalous ionization rate in the CMZ, but very close to it. 
%Yet, this is remarkable because the ionization rates expected from diffuse cosmic rays penetrating the CMZ are extremely suppressed and much lower than the values we find. Other powerful sources that are known to be within the CMZ has been investigated~\citep{Dogiel_2014, Ravikularaman2025AandA694A114} and can reach such high values only very near the source as well.

Our result rules out positrons from a variety of sources as the sole cause of the unexpectedly high ionization rate in the CMZ, at least within the assumption that positrons do not propagate significantly from their injections sources. However, we stress the high values of ionization expected in the inner parsec only by the positron injection that can be expected in the NSC.
Including diffusion and/or advection would slightly flatten the positron spatial distribution compared to what is shown in Fig.~\ref{fig:Distribs}. As a result, the ionization profile in the CMZ would become more uniform. Moreover, including diffuse reacceleration and advection can accelerate the injected positrons. This latter argument could reconcile the COMPTEL in-flight annihilation signal with the hypothesis of radionuclides producing the $511$~keV excess. However, the total level of ionization seems insufficient to reach the ionization rate measurements in any case and the effects of including propagation with winds, diffusion and reacceleration are expected to be very minor for MeV positrons.
%Furthermore, it could happen that positron propagation effects are more important than previously assumed, hence fattening the profile. However, even in that case, the total level of ionization seem insufficient to match the ionization rate measurements. %2) From the other side, although such a steep radial profile is not compatible with the observations, one can argue that positron diffusion and advection by winds could significantly flatten it, as discussed in~\cite{DelaTorreLuque:2024fcc}. Unfortunately, the lack of knowledge on the propagation conditions in this region make this possibility difficult to test with precision, and we leave it for a more dedicated study.

Most positron-producing sources, such as SNe and X-ray binaries, also emit other energetic particles that may significantly enhance the ionization and still provide a plausible explanation for this anomaly. Therefore, we still cannot exclude the possibility that the same sources producing the $511$~keV excess may be the responsible of the observed levels of ionization. Accounting for the ionization induced by accelerated protons and heavier nuclei is left for a future work. 
%Therefore, we conclude that explaining the anomalous CMZ ionization with the positrons responsible for the 511 keV line is disfavored. Nevertheless, it remains possible that the sources producing the 511 keV excess also drive the unexpectedly high CMZ ionization rate through the injection of other energetic particles.

Overall, this study highlights the potential correlation between excesses associated with low-energy particle emissions in the GC. Identifying such correlations could provide a powerful means of uncovering the origin of these anomalies and constraining the properties of previously unknown sources.%, which may also help to elucidate the long-standing \textit{Fermi}-LAT GC excess.

\section*{Acknowledgements}

The authors want to help Thomas Siegert and Sruthiranjani Ravikularaman for their useful and interesting comments and feedback.
PDL has been supported by the Juan de la Cierva JDC2022-048916-I grant, funded by MCIU/AEI/10.13039/501100011033 European Union "NextGenerationEU"/PRTR, and is currently supported by Ramón y Cajal RYC2024-048445-I grant, which is funded by MCIU/AEI/10.13039/501100011033 and FSE+. The work of PDL is also supported by the grants PID2021-125331NB-I00 and CEX2020-001007-S, both funded by MCIN/AEI/10.13039/501100011033 and by ``ERDF A way of making Europe''. PDL also acknowledges the MultiDark Network, ref. RED2022-134411-T. This project used computing resources from the National Academic Infrastructure for Supercomputing in Sweden (NAISS) under project NAISS 2024/5-666. The work of FC is supported by the Agence Nationale de la Recherche through the project ANR-25-CE31-6485.

%%%%%%%%%%%%%%%%%%%%%%%%%%%%%%%%%%%%%%%%%%%%%%%%%%
\section*{Data Availability}

The data used are fully public and appropriately referenced. All our results and scripts can be shared upon request.

%%%%%%%%%%%%%%%%%%%% REFERENCES %%%%%%%%%%%%%%%%%%

% The best way to enter references is to use BibTeX:

\bibliographystyle{mnras}
\bibliography{biblio} % 

%%%%%%%%%%%%%%%%%%%%%%%%%%%%%%%%%%%%%%%%%%%%%%%%%%

%%%%%%%%%%%%%%%%% APPENDICES %%%%%%%%%%%%%%%%%%%%%

%\newpage
%\onecolumn
\appendix
%\columnbreak

\section{Extra material}
\label{sec:App}
\begin{figure}
	\includegraphics[width=0.5\textwidth]{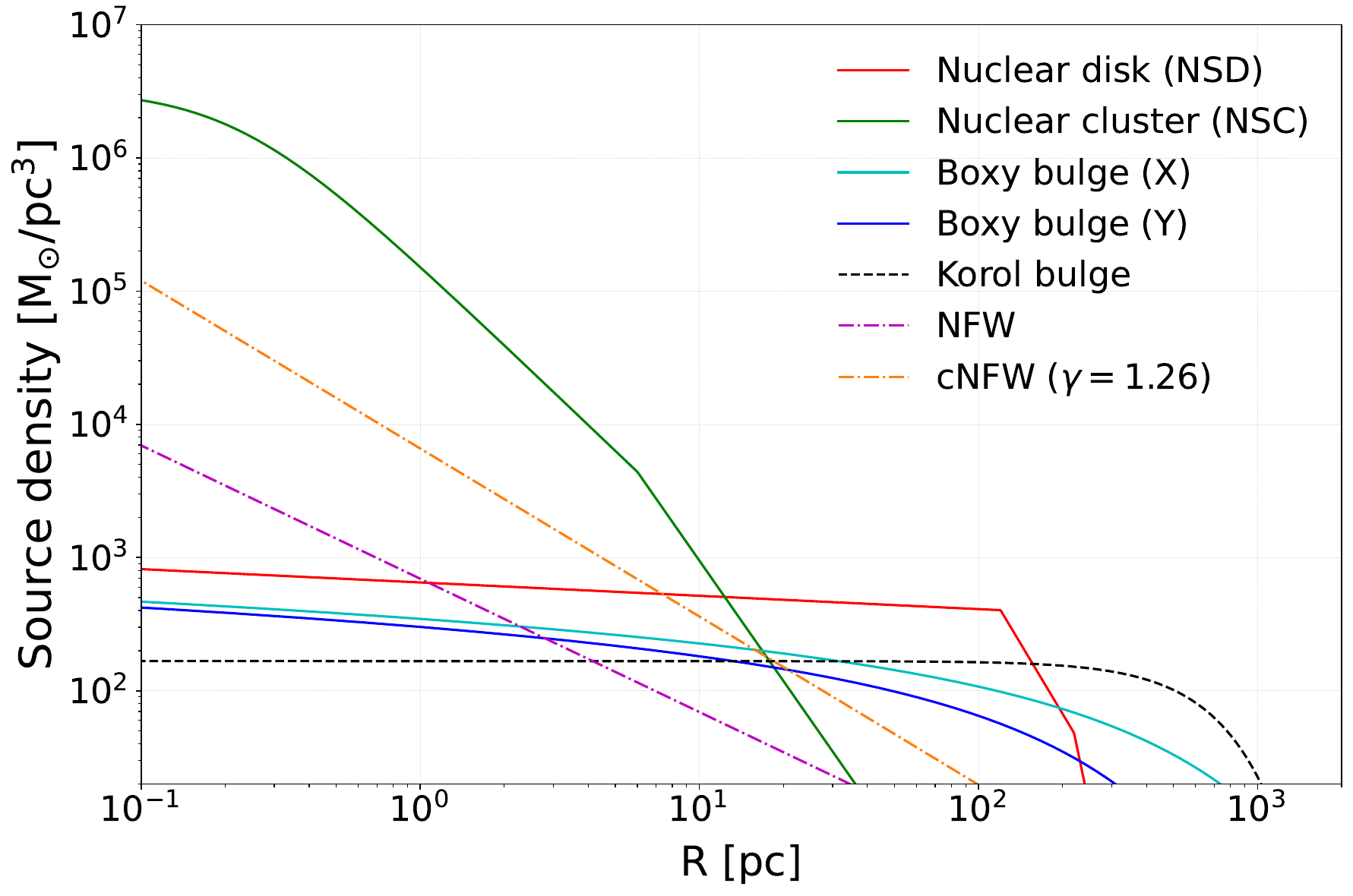}
    \includegraphics[width=0.5\textwidth]{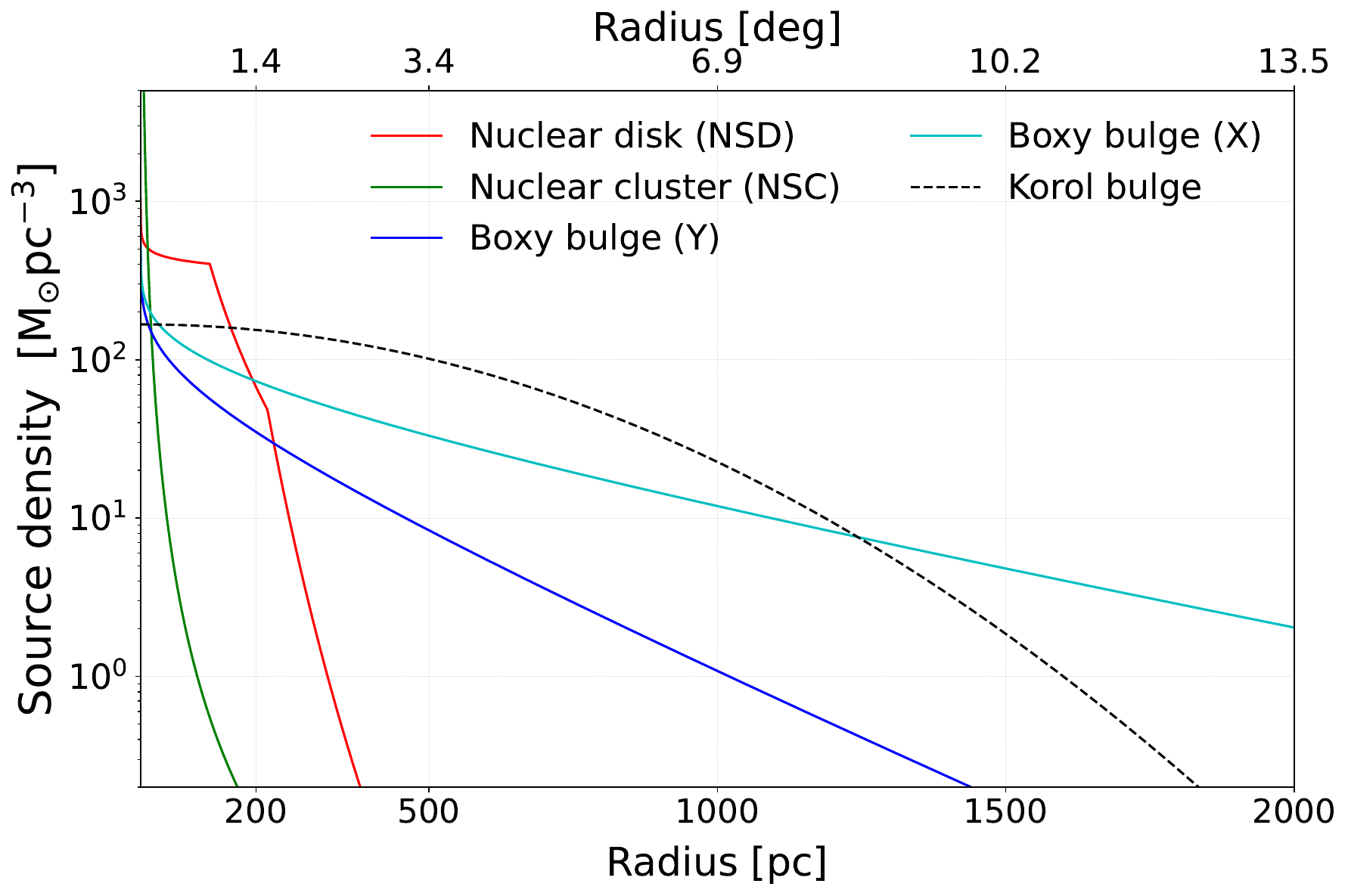}
    \caption{Same as shown in Fig.~\ref{fig:Distribs}, but including the dark matter density distribution expected from a Navarro-Frenk-White with inner slopes of 1 (classical NFW)  and 1.26 (referred to as cNFW in the legend).  The left panel focuses on the inner Galaxy, using a logarithmic x-axis, while the right panel uses a linear scale to highlight the large-scale behavior.}
    \label{fig:Distribs_wDM}
\end{figure}

In Fig.~\ref{fig:Distribs_wDM} we show a comparison of the different models for the distribution of sources at the bulge with the dark matter density distribution expected from a Navarro-Frenk-White parametrization and the same but with an inner slope (often also called contraction index) of 1.26 (similar to the one associated to the GCE). Both panels show the source density as a function of radial distance for each model, as in Fig.~\ref{fig:Distribs}.

\begin{figure*}
\includegraphics[width=1.2\columnwidth]{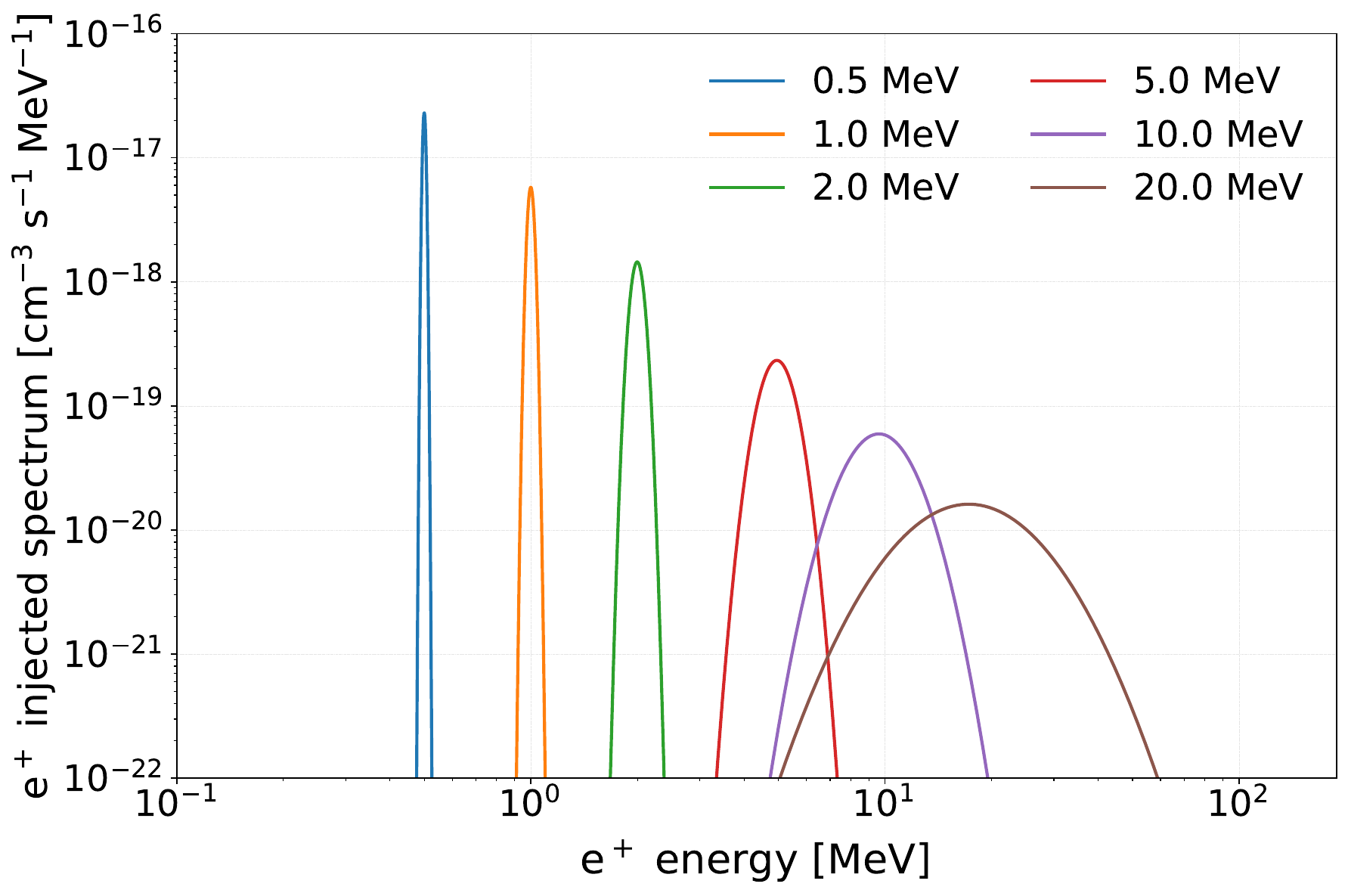}
\includegraphics[width=\columnwidth]{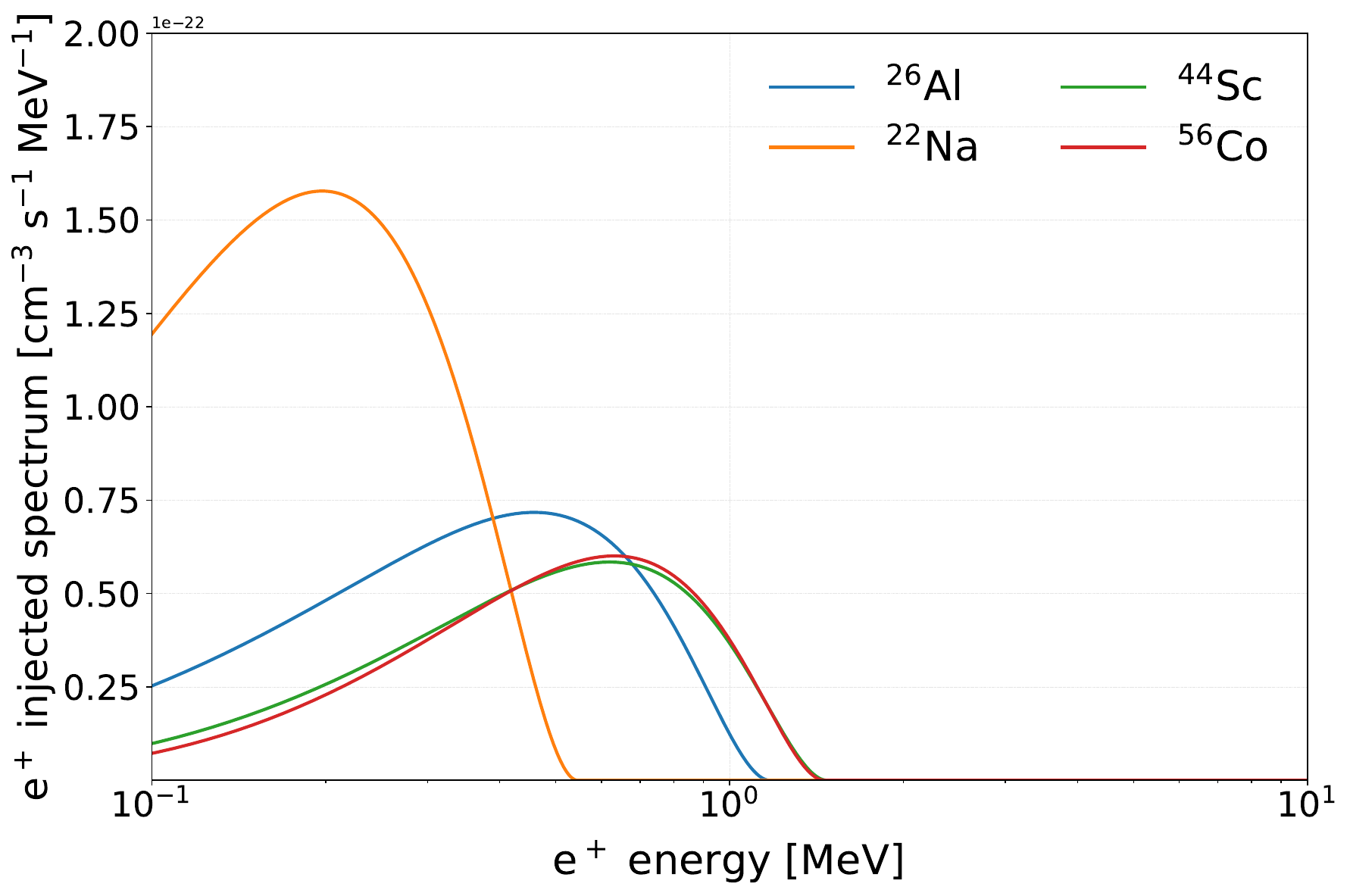}
\includegraphics[width=\columnwidth]{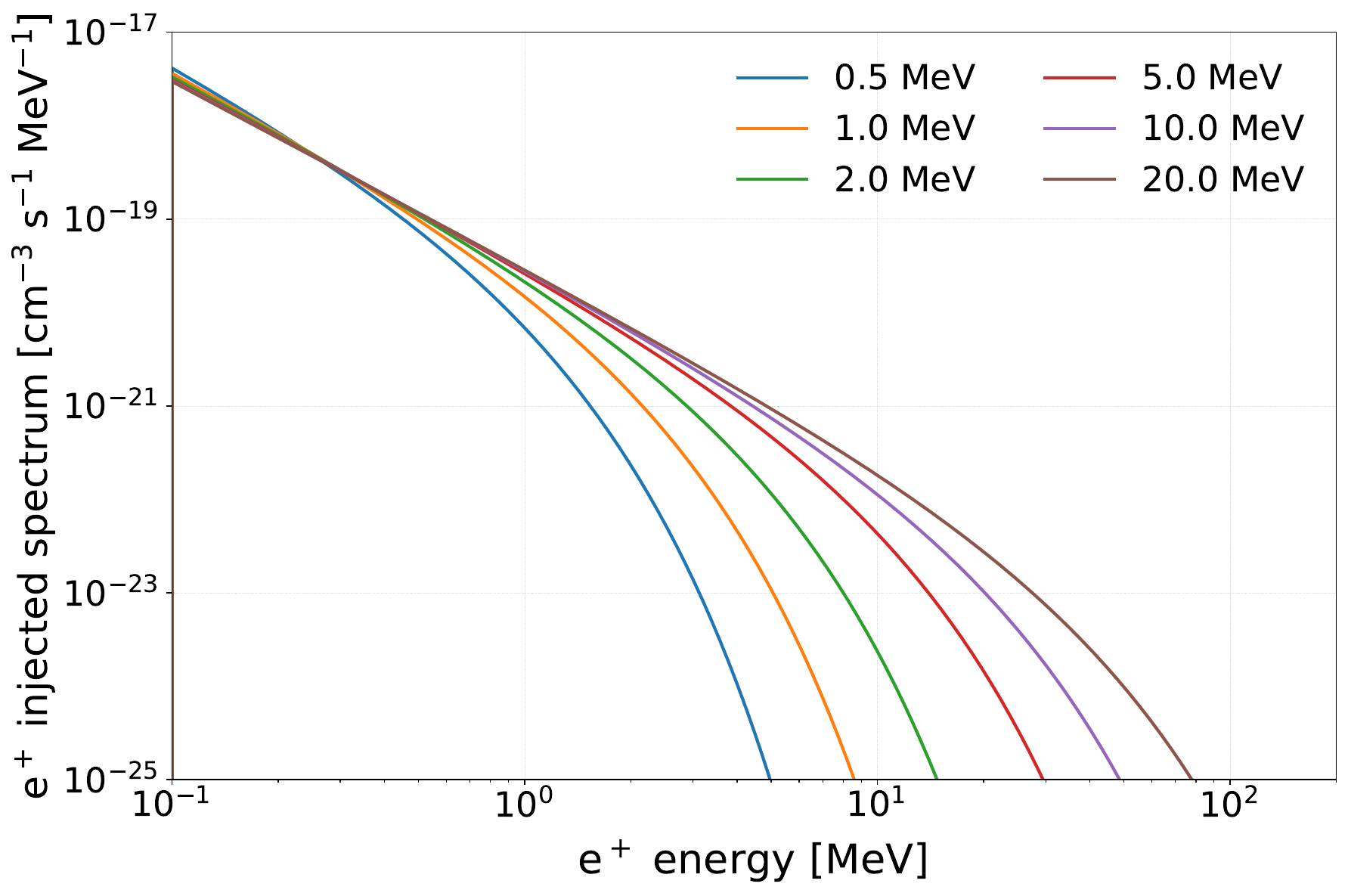}
    \caption{Positron injection spectra for the parametrizations used in this work. Here, we show the spectra normalized to a total positron injection rate of $2\cdot10^{43}$~e$^+$/s, adopting a distribution of sources following the spatial distribution of a combination of the nuclear stellar disk, nuclear stellar cluster and the boxy-bulge distributions.  
    The top panel refers to the case of assuming log-normal injection spectra, the left panel to the case of radionuclides injecting the positrons and the right panel to a power-law positron injection. In the top panel, the different colors refer to different E$_\text{mean}$ values, all for f=1/50. In the left panel, the different colors indicate the ionization rate from the radionuclides discussed in the main text. In the right panel, the different colors indicate different cut-off energies. In this panel, solid lines correspond to a power-law where the spectral index is $\gamma = 2.2$, while the dashed lines correspond to a power-law of $\gamma = 1.2$.}
    \label{fig:InjSpectra_2}
\end{figure*}

For completeness we also show the injection spectra used for the different parametrizations adopted in this work, and with the normalizations used (found from imposing a total positron injection rate that matches the $511$~keV observations), in Fig.~\ref{fig:InjSpectra_2}.
%\section{Extra material}
%\label{sec:App2}

\begin{figure*}
\includegraphics[width=\columnwidth]{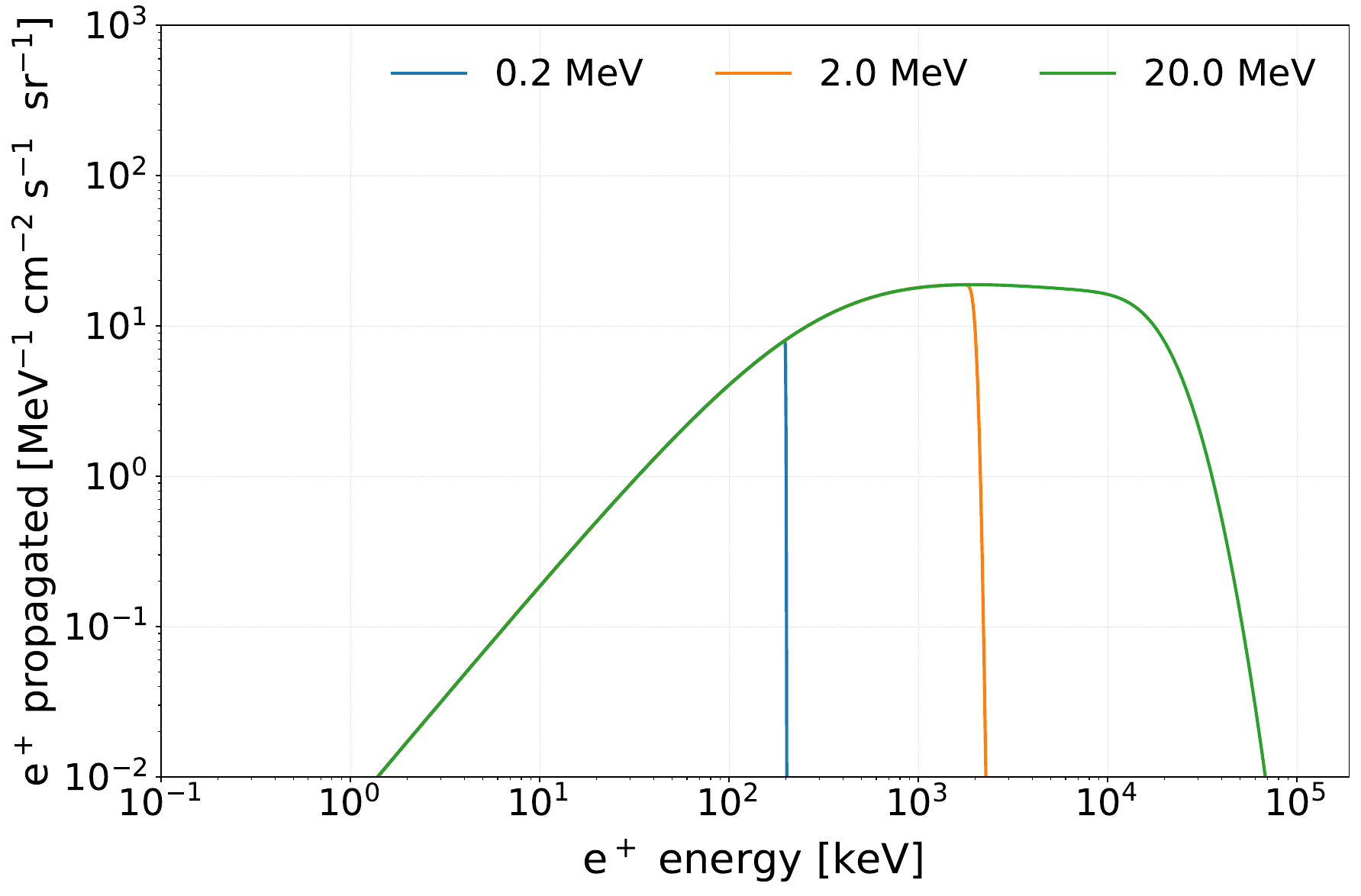}
\includegraphics[width=\columnwidth]{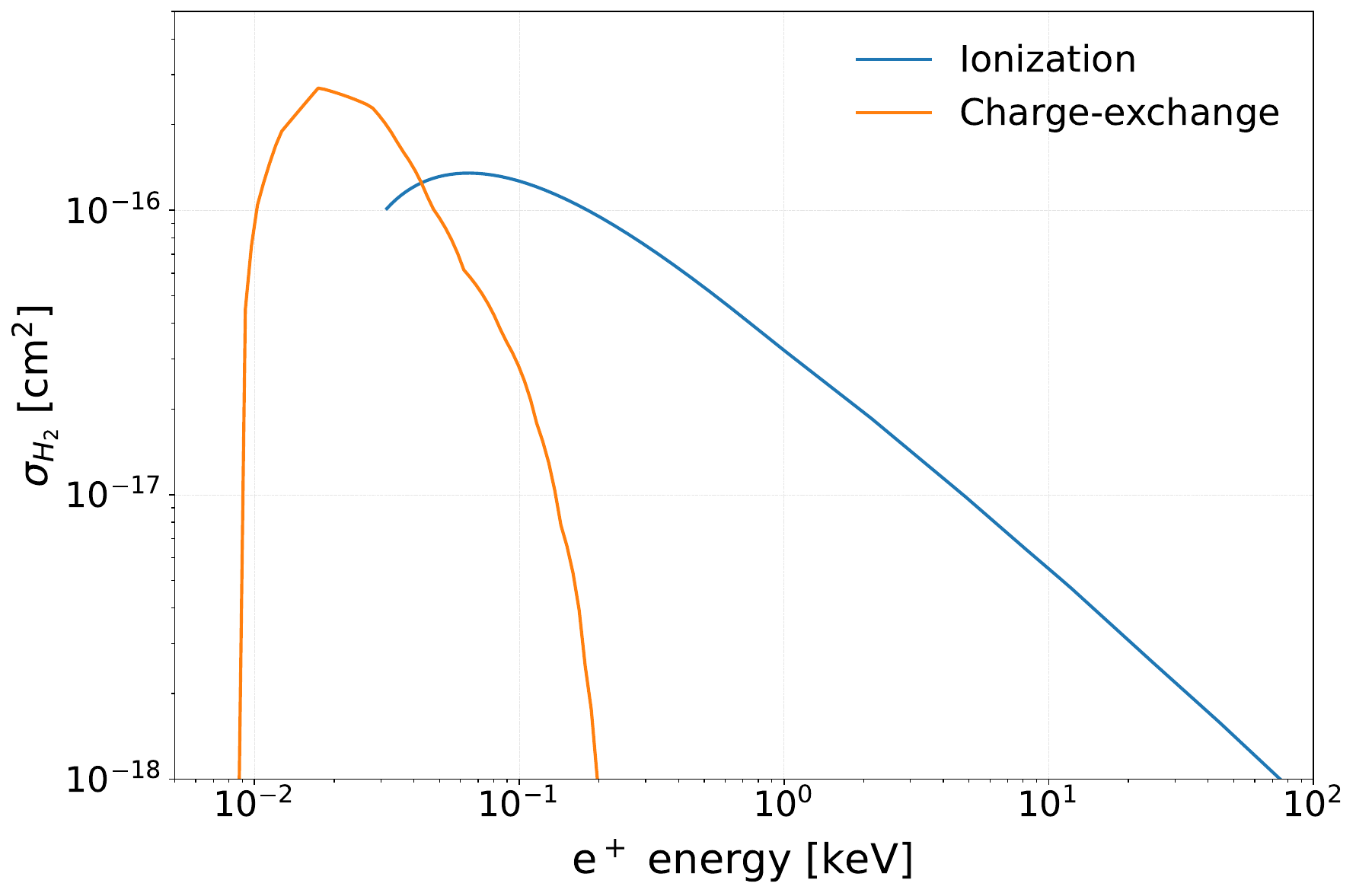}
    \caption{\textbf{Left panel:} Positron spectrum after energy losses for the general log-normal injection described above. The colors represent different mean injection energies of the log-normal distribution, as indicated in the legend. The spectra overlap at low energies because they depend on the energy integral (Eq.~\ref{eq:fluxe}), which is normalized to be the same for all signals (Eq.~\ref{eq:Norm}). \textbf{Right panel:} A comparison of the ionization cross sections of molecular hydrogen, with those of charge-exchange, which could potentially contribute to the CMZ ionization too.}
    \label{fig:ELoss_XS}
\end{figure*}

Additionally, we show in the left panel of Fig.~\ref{fig:ELoss_XS} the positron spectrum after energy losses for the general logarithmic injection described above. In the right panel of that figure, we show the ionization cross sections employed in this work, compared with those from charge-exchange with molecular hydrogen. Interestingly, both of them peak at a similar value of $\sim10^{-16}$~cm$^{-2}$. However, the charge-exchange cross sections are concentrated between 10-100 eV, whereas the ionization cross sections are still significant up to higher energies. Including the ionization produced from the charge-exchange process can be challenging, since it would mainly depend on the gas temperature (below the eV for a cold gas like that of the CMZ) and other details on how the positrons become thermal. If positrons thermalize below 10 eV, what can be expected for the cold gas in the CMZ, the charge-exchange process would lead to a negligible ionization in this region.

%%%%%%%%%%%%%%%%%%%%%%%%%%%%%%%%%%%%%%%%%%%%%%%%%%

% Don't change these lines
%\bsp	% typesetting comment
\label{lastpage}
\end{document}